\documentclass[%
reprint,
showpacs,preprintnumbers,
showkeys,
amsmath,amssymb,
aps,
pra,
]{revtex4-1}
\usepackage{tikz}
\usetikzlibrary{arrows,positioning,shapes,shadows,decorations.markings,decorations.pathmorphing}
\tikzset{->-/.style={decoration={markings,mark=at position #1 with {\arrow{>}}},postaction={decorate}}}
\tikzset{decay/.style={->,decorate,decoration={snake,amplitude=#1,segment length=1.5mm,post length=#1}}}
\usepackage{amsmath,amsfonts,amssymb,lineno,hyperref,appendix,url,ulem,cancel,wasysym,graphicx,sidecap,wrapfig,xcolor}
\usepackage{dcolumn}% Align table columns on decimal point
\usepackage{bm}% bold math
\newcommand{\ba}{\begin{equation}}
\newcommand{\bs}{\begin{split}}
\newcommand{\ea}{\end{equation}}
\newcommand{\baq}{\begin{eqnarray}}
\newcommand{\eaq}{\end{eqnarray}}

\newcommand{\eq}[1]{Eq.~(\ref{#1})}
\newcommand{\fig}[1]{Fig.~(\ref{#1})}

\newcommand{\app}[1]{Appendix.~(\ref{#1})}
\newcommand{\bra}{\langle}
\newcommand{\ket}{\rangle}
\newcommand{\td}{\tilde}
\newcommand{\+}{\dagger}
\newcommand{\dho}{\partial}
\newcommand{\dv}{\mathrm{d}}
\newcommand{\fd}[2]{\frac{\mathrm{d}#1}{\mathrm{d}#2}}
\newcommand{\pd}[2]{\frac{\partial #1}{\partial #2}}

\newcommand{\inti}{\int\limits_{-\infty}^{\infty}}
\newcommand{\nt}[2]{\int\limits_{#1}^{#2}}
\newcommand{\um}[2]{\sum\limits_{#1}^{#2}}
\newcommand{\g}{\gamma}

\newcommand{\de}{\delta}
\newcommand{\De}{\Delta}
\newcommand{\E}{\epsilon}

\newcommand{\e}{\eta}
\newcommand{\tx}{\theta}

\newcommand{\ka}{\kappa}

\newcommand{\s}{\sigma}

\newcommand{\q}{\phi}

\newcommand{\w}{\omega}
\newcommand{\W}{\Omega}
\newcommand{\x}{\chi}

\newcommand{\A}{\Lambda}

\begin{document}
	\preprint{APS/123-QED}
	\title{Quantum limit of hybrid atom-mechanical gyroscope based on electro-magnetically induced transparency}
	\author{Sankar Davuluri$^1$}
	\email{sd3964@csrc.ac.cn}
	\author{Yong Li$^{1,2}$}
	%\altaffiliation[Also at ]{Synergetic Innovation Center of Quantum Information and Quantum Physics, University of Science and Technology of China, Hefei 230026, China\\Synergetic Innovation Center for Quantum Effects and Applications, Hunan Normal University, Changsha 410081, China.}
	\affiliation{$^1$Beijing Computational Science Research Center, Beijing 100193, P. R. China}
	\affiliation{$^{2}$Synergetic Innovation Center for Quantum Effects and Applications, Hunan Normal University, Changsha 410081, China.}
	\date{\today}
	\begin{abstract}
		Application of hybrid atom-mechanical oscillator for absolute rotation detection is studied. The hybrid atom-mechanical oscillator consists of an atomic cell, filled with three level atoms, which is fixed on a mechanical oscillator. The atom-mechanical oscillator is placed on a rotating table such that the Coriolis force moves the atomic cell with respect to the incoming laser field. Thus the atomic resonance frequencies are Doppler shifted, and the phase of the laser field interacting with the atomic medium changes. Absolute rotation parameters are estimated by measuring the phase change in the laser field at the output of the atomic cell. Large dispersion is created in the atomic medium, using electromagnetically induced transparency, to enhance the phase change in the laser field interacting with the atomic medium. Contribution of the shot noise, the atomic noise and the noise due to the mechanical oscillation of the atomic cell are studied. We show that, under certain conditions, noise due to the mechanical oscillation of the atomic cell is on the same order of magnitude as the shot noise. The quantum limit of detectable rotation rate is estimated as $8.1\times10^{-19}$\,rad/s.
	\end{abstract}
	%Use showkeys class option if keyword
	\maketitle
	\section{Introduction}
	Detection of absolute rotation has significant importance in fundamental physics~\cite{everitt,stedman} and also in practical applications~\cite{smith,lee-is,ezekiel-alg}. The most prominent methods for rotation detection are fiber optic gyroscopes~\cite{arditty} (FOG), ring laser gyroscopes~\cite{chow} (RLG), matter wave interferometry~\cite{berg,gustavson} with Sagnac effect~\cite{malykin,kajari,arditty,chow}, and mechanical oscillators~\cite{sankar-njp,sankar-njp2,sankar-sc}. Here we propose an application of hybrid atom-mechanical oscillator for absolute rotation detection.
	
	The hybrid atom-mechanical oscillator consists of an atomic cell which is mounted on a mechanical oscillator. The atomic cell is filled with three level atoms to create slow-light\cite{hau,novikova} using electromagnetically induced transparency (EIT). The atom-mechanical oscillator is placed on a rotating table, and is driven co-sinusoidally along Y-axis. As a result, when the table rotates, the Coriolis force displaces the atomic cell along Z-axis. The motion of the atomic cell along Z-axis results in the Doppler shift in the atomic resonance frequencies with respect to the laser field which is propagating, along Z-axis, through the atomic cell. As a consequence, the phase of the laser field changes after interacting with the atomic medium. By measuring the phase change in the laser field at the output of the atomic cell, angular velocity of the rotating table can be estimated. In order to improve the phase sensitivity of the laser field interacting with atomic medium, we make the atomic medium dispersive~\cite{sautenkov}, but transparent, by using the EIT~\cite{harris,gray,sankar-ps,fleischhauer-rmp} phenomenon. Another advantage of working with EIT system is that its total quantum noise can set to be in the same order of magnitude as the  shot noise~\cite{fleischhauer-94} at the two-photon resonance for experimentally realizable parameters. Potential application of using EIT for rotation detection was described in \cite{sankar-epl1}, however a thorough investigation needs to be done.
	
	Coming to the mechanical oscillator part, recent advances in the field of optomechanics~\cite{aspelmeyer-rmp,meystre} demonstrate the design and control of high quality mechanical oscillators. We particularly consider a two-dimensional mechanical oscillator~\cite{norgia,faust-nat,faust-prl} which oscillates along Z-axis and Y-axis as shown in \fig{amg-f1}. The two-dimensional oscillator is driven co-sinusoidally along Y-axis. The frequency of the co-sinusoidal drive is chosen such that the Coriolis force acting on the oscillator, along Z-axis, is on resonance with the oscillator's frequency along Z-axis. This resonance condition ensures the optimal response of mechanical oscillator to the Coriolis force.
	\section{Principle}
	\begin{figure}[htb]
		\begin{tikzpicture}[scale=1,spring/.style={|-|,thick,decorate,decoration={coil,aspect=0.5,amplitude=1mm, segment length=1mm,post length=1mm, pre length=1mm}}] 
		%rotating platform
		\draw[black,line width=0.1mm,fill=black!10!](-40mm,-20mm)rectangle(40mm,20mm)node[]{};
		\node[rectangle,draw=black,fill=blue!10!,inner sep=2mm] (a) at (-30mm,0mm) {S};
		\node[rectangle,draw=black,fill=yellow!10!,inner sep=3mm,text width=10mm] (b) at (-0mm,0mm) {\hspace{2.5mm} M };
		\node[rectangle,draw=black,fill=brown!10!,inner sep=2mm] (c) at (30mm,0mm) {D};
		\draw[line width=0.3mm,blue,->-=0.3](a)->(b);
		\draw[line width=0.3mm,blue,->-=0.5](b)->(c);	
		\draw[spring,black](8mm,-3mm)-- (19mm,-3mm) node[]{};
		%        \draw[line width=1mm](a)--(-30mm,-10mm)--(30mm,-10mm)--(c);
		\draw[spring,black](0mm,-4mm)-- (0mm,-15mm) node[] {};
		\draw[->,line width=0.1mm](-37mm,10mm)arc(170:90:8mm)node[midway,above]{$\dot\tx$};
		\draw[->,line width=0.1mm] (25mm,10mm)--(25mm,15mm)node[above]{Y};
		\draw[->,line width=0.1mm](25mm,10mm)--(30mm,10mm)node[right]{Z};
		\end{tikzpicture}
		\caption{\label{amg-f2} A Laser source `S' and a phase sensitive photo-detector `D' are rigidly fixed with respect to each other on a table rotating with angular velocity $\dot\tx$. An optical medium `M' with frequency dependent refractive index $\x$ is mounted on a two-dimensional mechanical oscillator which is fixed to the rotating table. A classical electromagnetic field (shown with blue arrow) with frequency $\w'_o$ passes through M.}
	\end{figure}
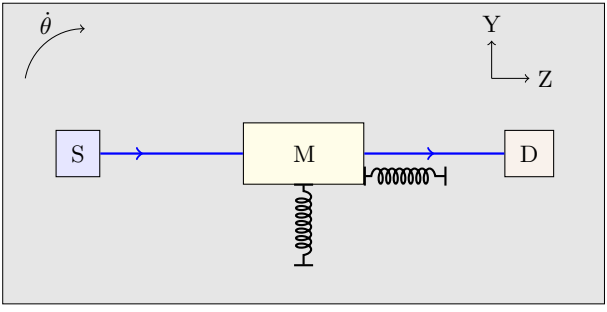
	For illustration purpose, assume that $\chi(\w')$ is the frequency-dependent refractive index of the atomic medium mounted on a two-dimensional mechanical oscillator shown in \fig{amg-f2}. The laser source S, detector D, and the two-dimensional mechanical oscillator is fixed on a table (shown by grey rectangle in \fig{amg-f2})rotating with angular velocity $\dot\tx'$. When $\dot\tx'=0$, change in the phase $\q'_1$ of the classical laser field after passing through the atomic medium is given as
	\ba
	\q'_1=\x(\w'_o)L'\frac{\w'_o}{c}, \label{amg-11}
	\ea
	where $\w'_o/2\pi$ is the frequency of the classical laser field, $L'$ is the length of the atomic cell. The oscillator is driven co-sinusoidally along Y-axis at a frequency equal to the resonance frequency of the oscillator along Z-axis. As a result, when the table rotates with angular velocity $\dot\tx'$, the Coriolis force acting on the mechanical oscillator  moves the atomic cell along Z-axis. Hence the laser frequency is Doppler shifted by $\w''$ with respect to the atomic resonance frequency. So when the table is rotating, change in the phase $\q'_2$ of the laser field after passing through the moving atomic medium is given as
	\ba\bs
	\q_2&=\frac{L}{c}\x(\w'_o+\w'')(\w'_o+\w'')\\&=\frac{L'}{c}[\x(\w'_o)\w'_o+\w''(\x(\w'_o)+\w'_o\fd{\x}{\w'}|_{\w'=\w'_o})+\cdots]. \label{amg-16}
\end{split}\ea
Assuming that non-linear dispersion in \eq{amg-16} is negligible, change in the phase of the laser field due to absolute rotation can be obtained from \eq{amg-11} and \eq{amg-16} as
\ba
\q'_2-\q'_1=\frac{L'}{c}[\w''(\x(\w'_o)+\w'_o\fd{\x}{\w'}|_{\w'=\w'_o})].\label{amg-14}
\ea
Equation~(\ref{amg-14}) shows that a larger linear dispersion of the atomic medium will enable us to measure smaller frequency shift $\w''$. Since $\w''$ is a consequence of rotation, we can estimate the angular velocity of rotation from $\w''$. Based on the idea presented in this section, we propose a hybrid atom-mechanical gyroscope in the next section.
\section{Equations of motion}
\begin{figure*}[htb]
	\begin{tikzpicture}[scale=1,spring/.style={|-|,thick,decorate,decoration={coil,aspect=0.8,amplitude=1mm, segment length=1.3mm,post length=1mm, pre length=1mm}}] 
	%Rotating platform
	\node[rectangle,draw=black,fill=black!10!,minimum width=4mm,minimum height=1mm,scale=35]at (1.1,0.4){};
	% Draw the arc with center is (2,1)
	\draw[thick,->] (-5,3) arc (160:110:1.5cm)node[midway,above]{$\dot\tx$};
	%Mechanical drive
	\draw[draw=black, rounded corners,fill=brown] (-0.9,-1.6)rectangle(1.3,-0.15)node[]{};
		\draw[spring] (0,-1.6) -- (0,-3.25) node[] at (1.5,-2.5) {$\dot{\bar y}=\dot{\bar y}_o\cos(2\pi\nu t)$};
	\draw[<->,thick](.2,-1.8)--(.2,-3.)node[]{};
	%detectors and laser
	\node[rectangle,draw=black,line width=0.3mm,fill=green!10!,inner sep=2.5mm, minimum size=2.5] (b) at (-4,-1) {LS};
	\node[rectangle,draw=black,fill=green!10!,inner sep=2.5, line width=1](c) at (6,2) {D1};
	\node[rectangle,draw=black,fill=green!10!,inner sep=2.5, line width=1] (d)at (5,3) {D2};
	%Interferometer laser path
	\draw[line width=2mm,blue!80!](b)--(5,-1)node[above]at(-1.5,-1){arm-1};
	\draw[line width=2mm,blue!80!](5,-1)--(d)node[]{};
	\draw[line width=2mm,blue!80!](-3,-1)--(-3,2)node[]{};
	\draw[line width=2mm,blue!80!](-3,2)--(c)node[above]at(-1,2){arm-2};
	%drive laser
	\draw[red,fill=red,opacity=0.5](0,-1) circle (0.25)node[]{};
	%Beam splitters and mirrors
	\draw[line width=4, blue!50!](-3.2,-1.2)--(-2.8,-0.8)node[]{};
	\draw[line width=4, blue!50!](4.8,1.8)--(5.2,2.2)node[]{};
	\draw[line width=4](4.8,-1.2)--(5.2,-0.8)node[]{};
	\draw[line width=4](-2.8,2.2)--(-3.2,1.8)node[]{};
	%arm-1
	%atom-mechanical oscillator
	\draw[green,fill=green,opacity=0.4](-.1,-1.4)rectangle(0.1,-0.6)node[]{};
	\draw[spring,black!10!green] (0.1,-1.3) -- (1,-1.3) node[midway,right] {};
	%atoms
	\draw[yellow,fill=yellow](-0.05,-1.03)rectangle(0.05, -0.97)node[]{};
	%\draw[thick,dotted] (0,-.97)--(2.3,-0.1) node[]{};
	%atomic scheme
	\node[circle,draw=yellow,fill=yellow!10!,minimum size=60] at (3.4,0.3){};
	\draw[thick] (3.5,-0.2) -- (4,-0.2) node[below,midway] {$|b\ket$};
	\draw[thick] (3,1) -- (3.5,1) node[right] {$|a\ket$};
	\draw[thick] (2.5,0.2) -- (3,0.2) node[midway,below] {$|c\ket$};
	\draw[thin,blue,<->,>=stealth] (3.75,-0.2) -- (3.27,1) node[midway,right] {$\hat a$};
	\draw[thick,red,<->,>=stealth] (2.75,0.2) -- (3.2,1) node[midway,left] {$\W $};
	%arm-2
	\node[rectangle, draw=black,fill=green!10!,minimum size=10,line width=1,text width=25]at(1,2){$\pi/2$\\Phase shifter};
	% axis
	\draw[thick,->] (6,-3)--(7,-3)node[midway,below]{Z-axis};
	\draw[thick,->] (6,-3)--(6,-2)node[midway,above,rotate=90]{Y-axis};
	%\draw [step=1.0,blue, very thick] (-6,-3) grid (6,4);
	\end{tikzpicture}
	\caption{LS: Laser source, D1,D2: photo detector. The small yellow rectangle represents cold (temperature of atoms $T_a=10^{-6}$\ K) atomic cell which is fixed on a mechanical oscillator shown in green color. We call this set-up of atomic-cell fixed on the mechanical oscillator as atom-mechanical oscillator. The atom-mechanical oscillator can oscillate along Z-axis and its resonance frequency is $\nu$. The brown rectangle represent the driving platform on which the atom-mechanical oscillator is fixed. The driving platform is driven along Y-axis with velocity $\dot{\bar y}=\dot{\bar y}_o\cos(2\pi\nu t)$, where $t$ is time and $\dot{\bar y}_o$ is maximum mean velocity along Y-axis. The energy level structure of the atoms inside the atomic cell is shown in big yellow circle. The red circle represents the cross-section area of the drive laser which is coupling $|a\ket-|c\ket$ transition and is propagating perpendicular to the YZ-plane. The blue laser field in the arm-1 of the interferometer is the weak quantum probe laser field which couples the $|a\ket-|b\ket$ transition while propagating along Z-axis in arm-1. We assume that the diameter $L_d$ of probe and drive laser beams is larger than $\dot{\bar y}_o/\nu$, so that the number of atoms in the laser-atom interaction region do not change because of the driving platform's oscillation along Y-axis. The complete set-up is fixed on a platform (the big grey rectangle) rotating with angular velocity $\dot\tx$. The Z-axis and Y-axis shown in the figure are fixed with respect to the rotating platform. When $\dot\tx\neq0$ the atom-mechanical cell moves along the Z-axis with velocity $\dot{\hat z}_m$, $\hat z_m$ is center of mass of the atom-mechanical oscillator, because of the Coriolis force acting on it. As a result the resonance frequency of the atoms in the atom-mechanical oscillator are Doppler shifted with respect to the probe laser. This rotation induced Doppler frequency shift leads to phase shift which is measured at D1 and D2 to estimate $\dot\tx$.}
	\label{amg-f1}
\end{figure*}
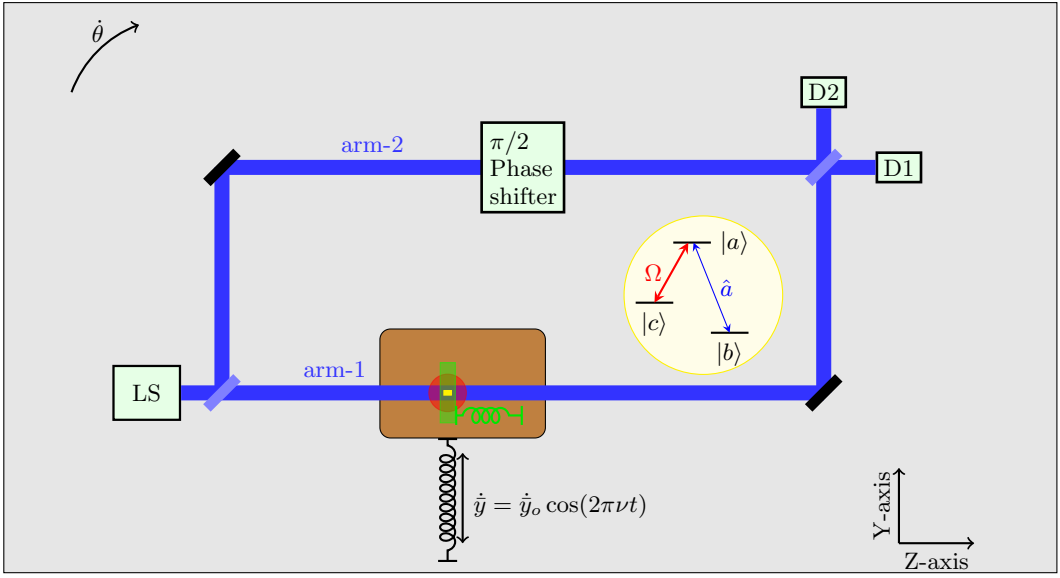
Schematics of the atom-mechanical gyroscope are as shown in \fig{amg-f1}. An atomic-cell (shown by yellow rectangle in \fig{amg-f1}) of length $L$ is placed in the arm-1 of the Mach-Zehnder interferometer. The atomic cell is fixed on a mechanical oscillator (shown in green color in \fig{amg-f1}) and we call this set-up as atom-mechanical oscillator. The atom-mechanical oscillator can oscillate along Z-axis with resonance frequency $\nu$. The atom-mechanical oscillator is fixed on another platform (shown in brown color in \fig{amg-f1}), we call this platform as driving platform, which is driven along Y-axis with mean velocity $\dot{\bar y}=\dot{\bar y}_o\cos(2\pi\nu t)$. Where $t$ is time and $\dot{\bar y}_o$ is the maximum mean velocity of the driving platform.  This complete set-up is placed on a platform (shown in grey color in \fig{amg-f1}) rotating with angular velocity $\dot\tx$.
	
	Atoms inside the atomic cell are three-level atoms with energy of each atomic level $|u\ket (u=a,b,c)$ given by $\hbar \w_u$, where $\hbar$ is the reduced Planck constant and $\w_u/2\pi$ is the frequency. The energy level structure of the atoms inside the atomic cell is shown in big yellow circle is \fig{amg-f1}. Transition $|b\ket -|c\ket$ is electric-dipole forbidden. Transition $|a\ket-|b\ket$, with electric-dipole moment $\wp_{ab}$, is coupled to the weak quantum probe field (shown in blue color in \fig{amg-f1}) propagating along Z-axis in the arm-1 of the interferometer. Transition $|a\ket-|c\ket$, with dipole moment $\wp_{ac}$, is driven by a strong classical laser field for which the cross-section area is shown by the red circle in \fig{amg-f2}.  The classical drive field has frequency $\omega_d/2\pi$ and is propagating perpendicular to the YZ-plane (see \fig{amg-f1}). We assume that the diameter $L_d$ of both probe and drive laser fields is larger than $\dot{\bar y}_o/\nu$, so that the number of atoms in the laser-atom interaction region do not change because of the driving platform's oscillation along Y-axis. This condition can easily be satisfied for realistic parameters like $L_d=2\times10^{-2}$\,m, $\dot{\bar y}_o=10^{-2}$\,m/s and $\nu=1$\,Hz.

By setting $\dot{\tx}=0$ and $\dot{\bar y}=0$, propagation~\cite{abram} of quantum electromagnetic field is described by using multi-mode description~\cite{blow}. Hence we consider the quantum probe field coupling with $|a\ket-|b\ket$ transition as the quasi-monochromatic field, represented by $\um{r}{}\hat c_r$, with mean wave-vector $k_p$. The wave vector for the $r$-th mode of the probe field is
\ba
k_r=k_p+\frac{2r\pi}{L_q},\quad\w_r=\w_p+\frac{2r\pi c}{L_q}, \quad r=-R,...,R,
\ea
where $c$ is the velocity of light in vacuum, $\omega_r=ck_r$, $\omega_p=ck_p$, and $L_q$ is the quantization length so that the quantization volume is $AL_q$ with $A$ as the area of cross section of electromagnetic field to be quantized. Interaction between the quantum field and the atoms in a sample of length $L$ can be described by using the technique described in Refs.~\cite{drummond, fleischhauer-95, fleischhauer-94, das}. So we divide the total interaction length $L$ into $(2P+1)$ sub-cells. Each sub-cell is centered at $z_l=lL/(2P+1), l=-P,...,P$, such that $\De z:=z_{l+1}-z_l=L/(2P+1).$ Single-atom operators are represented by $\hat\s^{ls}_{uv}=|u\ket ^{ls } \bra v|^{ls}$, where $v=a,b,c$, superscripts $s$ and $l$ denote the $s$-th atom in the $l$-th sub-cell centered at $z_l$. Hence the Hamiltonian for atom-field interaction, when $\dot\tx=0$ and $\dot{\bar y}=0$, is given as
\ba\bs
\hat H=\big(\hbar\um{s,l,r}{}g_r\hat\s_{ab}^{ls}\hat c_re^{i(k_r-k_p)z_l}+\um{l,s}{}\hbar\td\Omega\hat\s_{ac}^{ls}+\\
+HC\big)+\um{r}{}\hbar\w_{r}(\hat c_r^\+\hat c_r+\frac{1}{2})+\um{l,s,u}{}\hbar\w_u\hat\s^{ls}_{uu},\label{amg-ham}
\end{split}\ea
where HC is the Hermitian conjugate, $\td\W=-\wp_{ac}E_de^{-i\w_d t}/\hbar$ with $E_d$ being the amplitude of the drive field,  $g_r=-(\wp_{ab}/\hbar)\sqrt{\hbar \w_r/\E_oAL_q}.$ In the limit $L_q\to\infty$ and $\De z\to 0$, by defining
\begin{subequations}
\ba\bs
\frac{2R+1}{N}\um{s}{}\hat\s_{uv}^{ls}|_{z_l\to z}:=\hat\s_{ij}^o,\\
\frac{2R+1}{N}\um{s}{}\hat F_{uv}^{ls}|_{z_l\to z}:=\hat F^o_{ij},\\
\frac{1}{\sqrt{2\pi}}\nt{0}{\infty}\dv\w \hat c(\w)e^{-i\w t}e^{i(k-k_p)z_l}|_{z_l\to z}:=\hat ae^{-i\w_pt},
\end{split}\ea\label{amg-dc}
\end{subequations}
equations of motion (refer \app{amg-a1} for more details) for the quantum field and the atomic operators are given as
\begin{subequations}
\ba
\Big(\frac{\dho}{\dho t}+c\frac{\dho}{\dho z}\Big)\hat a=-iNg_c^*\frac{c}{L_q}\hat\s_{ba},
\ea
\ba
\dot{\hat\s}_{ba}=[i(-\w_{ab}+\w_p)-\g_{ab}]\hat\s_{ba}+ig_c(\hat\s_{aa}-\hat\s_{bb})\hat a-i\W\hat\s_{bc}+\hat F_{ba},
\ea
\ba
\dot {\hat\s}_{bc}=[i(-\w_{cb}+(\w_p-\w_d))-\g_{bc}]\hat\s_{bc}+ig_c\hat\s_{ac}\hat a-i\W\hat\s_{ba}+\hat F_{bc},\label{amg-7-1}
\ea
\ba
\dot{\hat\s}_{ac}=[i(\w_{ac}-\w_d)-\g_{ac}] \hat\s_{ac}+i(\hat\s_{cc}-\hat\s_{aa})\W+ig_c^*\hat\s_{bc}\hat a^{\+}+\hat F_{ac},
\ea
\label{amg-7}
\end{subequations}
where $g_c=-\wp_{ab}/\hbar(\sqrt{\hbar\w_p/\E_oAc}),$ $\w_{uv}=\w_u-\w_v$, $\W=-\wp_{ac}E_d/\hbar $ is taken as a real quantity, $\g_{uv}$ is the de-coherence between the atomic levels $|u\ket$ and $|v\ket$, $\hat\s_{uv}$ and $\hat F_{uv}$ are the rotating wave approximated operators of $\hat\s_{uv}^o$ and $\hat F_{uv}^o$, respectievely. Even though the transition $|b\ket-|c\ket$ is electric dipole forbidden, other factors like line-width of the drive laser lead to decoherence on $|b\ket-|c\ket$ transition. Hence we assumed phenomenological $\g_{bc}$ in \eq{amg-7-1}.

For $\dot\tx\neq0$ and $\dot{\bar y}=\dot{\bar y}_o\cos(2\pi\nu t)$, a classical Coriolis force of $2m\dot{\bar y}_o\dot\tx\cos(2\pi\nu t)$ is exerted on the atom-mechanical oscillator cell, along the Z-axis, when the table rotates with angular velocity $\dot\tx$. Equation of motion of the atom-mechanical oscillator along Z-axis is given as
\ba
\ddot{\hat z}_m+\frac{\nu}{Q}\dot{\hat z}_m+(2\pi\nu)^2 \hat z_m=2\dot{\bar y}_o\dot{\tx}\cos(2\pi\nu t)+\hat F_{th},\label{amg-1}
\ea
where $Q$ and $\hat z_m$ are the quality factor and center of mass position of the atom-mechanical oscillator, respectively. $\hat F_{th}$ is the thermal noise operator. Substituting the ansatz $\hat z_m=\hat z_{mo}\cos(2\pi\nu t)$ in \eq{amg-1} gives
\ba
\dot{\hat z}_m=\frac{2Q\dot{\bar y}_o\dot{\tx}}{\nu}\cos(2\pi\nu t)+\frac{Q}{\nu}\hat F_{th}.\label{amg-vel}
\ea
As the atomic cell is moving with velocity $\dot{\hat z}_m$ with respect to the laser fields, resonance frequencies of the atomic energy levels are shifted due to the Doppler effect as
\ba
\w_{ab}\to\w_{ab}(1-\frac{\dot{\hat z}_m}{c}).\label{amg-2}
\ea
By substituting \eq{amg-2} in \eq{amg-7}, and by setting $\w_p=\w_{ab}$ and $\w_{d}=\w_{ac}$, we can write
\begin{subequations}
\ba
\Big(\frac{\dho}{\dho t}+c\frac{\dho}{\dho z}\Big)\hat a=-iNg_c^*\frac{c}{L_q}\hat\s_{ba},
\ea
\ba
\dot{\hat\s}_{ba}=(i\w_{ab}\frac{\dot{\hat z}_m}{c}-\g_{ab})\hat\s_{ba}+ig_c(\hat\s_{aa}-\hat\s_{bb})a-i\W\hat\s_{bc}+\hat F_{ba},
\ea
\ba
\dot{\hat\s}_{bc}=(i\w_{ab}\frac{\dot{\hat z}_m}{c}-\g_{bc})\hat\s_{bc}+ig_c\hat\s_{ac}a-i\W\hat\s_{ba}+\hat F_{bc},
\ea
\ba
\dot{\hat\s}_{ac}=(-\g_{ac})\hat\s_{ac}+i(\hat\s_{cc}-\hat\s_{aa})\W+ig_c^*\hat\s_{bc}\hat a^{\+}+\hat F_{ac}.
\ea
\label{amg-we}
\end{subequations}
Note that the $\dot{\hat z}_m$ in \eq{amg-we} has the information about $\dot\tx$ as given in \eq{amg-vel}.
\subsection{Signal}
Equations~(\ref{amg-we}) are linearized by writing operators as $\hat \s_{uv}=\bar\s_{uv}+\hat\de_{uv},$ and $\hat a=\bar a+\hat\de,$ with $\bar\s_{uv}$, $\hat{\de}_{uv}$, $\bar a$, and $\hat \de$ as the classical mean of $\hat{\s}_{uv}$, quantum fluctuation in $\hat\s_{uv},$ classical mean of $\hat a$ and quantum fluctuation in $\hat a$, respectively. Vacuum field entering arm-1 is included in $\hat\de$. Assuming EIT conditions, we treat $\hat a$ only up to its first order while $\W$ is kept to all orders. Most of the atomic population stays in $|b\ket$ under EIT conditions. Hence, we can write~\cite{dantan-04,sun} $\hat\s^0_{bb}=1$ and $\hat\s^0_{aa}=\hat\s^0_{cc}=\hat\s^0_{ac}=0$(the superscript `0' indicates zeroth order in $\hat a$, while the superscript `1' indicates first order in $\hat a$). For realistic parameters: $\g_{ab}\approx\g_{ac}\approx 10^{7}$\,Hz, while $\nu\approx 1$Hz. Because $\nu\ll\g_{ab}\approx\g_{ac}$, by using adiabatic approximation\cite{marlan-text}, $\bar\s_{ba}^1$ is evaluated as
\ba
\bar\s^1_{ba}=\frac{ig_c\bar a(i\w_{ab}\frac{\dot{\bar z}_m}{c}-\g_{bc})}{(i\w_{ab}\frac{\dot{\bar z}_m}{c}-\g_{ab})(i\w_{ab}\frac{\dot{\bar z}_m}{c}-\g_{bc})+\W^2},\label{amg-sba}
\ea
where $\dot{\bar z}_m=2\dot{\bar y}_o\dot\tx\cos(2\pi\nu t)Q/\nu.$ Using \eq{amg-sba}, the propagation of steady state probe field $\bar a$ is given\cite{fleischhauer-00} as
\ba
\frac{\dho}{\dho z}\bar a=\A_o\bar a,\label{amg-mxs}
\ea
where $$ \A_o= \ka\frac{i\w_{ab}\frac{\dot{\bar z}_m}{c}-\g_{bc}}{(i\w_{ab}\frac{\dot{\bar z}_m}{c}-\g_{ab})(i\w_{ab}\frac{\dot{\bar z}_m}{c}-\g_{bc})+\W^2},$$ with $\ka=N|g|^2/c,$ and $g=g_c\sqrt{c/L_q}$. Solving \eq{amg-mxs} gives
\ba
\bar a(L)=\bar a_oe^{\A_oL},\label{amg-8}
\ea
where $\bar a_o=\bar a(z=0)$. Assuming that $\W^2\gg\g_{ab}\g_{bc}$, and by considering $\w_{cb}\dot{\bar z}_m/c$ and $\w_{ab}\dot{\bar z}_m/c$ upto their first order only, $\A_o$ can be approximated as
\ba
\A_o\approx \ka\Big(-\frac{\g_{bc}}{\W^2}+i\frac{\w_{ab}\frac{\dot{\bar z}_m}{c}}{\W^2}\Big).\label{amg-tay}
\ea
Electromagnetic field in the arm-2 of \fig{amg-f1} is represented by $\hat a_1$. Hence the difference in the photo detector readings, after adding a constant $\pi/2$ phase\cite{caves} to $\hat a_1$, is given as
\ba
\hat I_1-\hat I_2=\hat a^\+\hat a_1+\hat a\hat a_1^\+.\label{amg-h}
\ea
The mean difference in the photo-detector readings is given as
\ba
\bra\hat I_1-\hat I_2\ket=\bar a^*\bar a_1+\bar a\bar a_1^*,\label{amg-hs}
\ea
where $\bar a_1=\hat a_1-\hat\de_1$ with $\hat\de_1$ as quantum fluctuation in $\hat a_1$. Vacuum field entering arm-2 is included in $\hat\de_1$. By using the relation $\bar a(z_o)=-i\bar a_1$ and using \eq{amg-vel}, \eq{amg-tay}, and \eq{amg-8}, we can write \eq{amg-hs} as
\ba\bs
\bra I_1-I_2\ket\approx-\frac{2|\bar a|^22\dot{\bar y}_o\dot{\tx}}{\nu/Q}e^{-\frac{\ka\g_{bc}}{\W^2}L}(\frac{\ka\w_{ab}}{c\W^2}L)\cos(2\pi\nu t).\label{amg-10}
\end{split}\ea
Minimizing \eq{amg-10} with respect to $\ka\g_{bc}L/\W^2$ shows that we obtain maximum signal when $\ka\g_{bc}L=\W^2$. At $\W^2=\ka\g_{bc}L$, signal is given as
\ba
\bra I_1-I_2\ket=-\frac{4|\bar a|^2\dot{\bar y}_o\dot{\tx}}{\nu/Q}e^{-1}(\frac{\w_{ab}}{c\g_{bc}})\cos(2\pi\nu t).\label{amg-os}
\ea
Signal in \eq{amg-os} represents the number of photons detected per unit time. Hence the total number of photons detected in a measurement time of $t_m$ is given as
\ba
\nt{t_o}{t_o+t_m}\bra I_1-I_2\ket \dv t.\label{amg-19}
\ea
Assuming that $t_m\ll 1/\nu$ and considering that $t_o=0$, we can simplify \eq{amg-19} as
\ba
S_o:=\nt{0}{t_m}\bra I_1-I_2\ket dt=\frac{4\dot{\bar y}_o\dot{\tx}\w_{ab}}{c\g_{bc}\nu/Q}\frac{Pt_m}{\hbar\w_p}e^{-1},\label{amg-sig}
\ea
where $P$ is the power of the probe field. Since $t_m\ll1/\nu$, there is no significant change in velocity of the atomic cell during the time of measurement. Hence, the system behaves as if atomic cell is moving with constant velocity, which is equal to the velocity at $t_o$, with respect to the incoming electromagnetic field. For $t_m\ll1/\nu$, it is enough to consider the effect of $\nu$ up to its first order.
\section{Noise spectrum}
Shot noise, atomic noise, and mechanical noise of the atom-mechanical oscillator are the three major sources of noise in the hybrid atom-mechanical system. Noise from the mechanical motion of the atomic cell is given by $\hat F_{th}$ term in \eq{amg-vel}. By substituting \eq{amg-vel} in \eq{amg-we}, the linearized equations of motion for the fluctuations are
\begin{subequations}
\ba
\Big(\frac{\dho}{\dho t}+c\frac{\dho}{\dho z}\Big)\hat \de=-iNg_c^*\frac{c}{L_q}\hat\de^1_{ba},\label{amg-mx}
\ea
\ba
\dot{\hat{\de}}^1_{ba}=(i\w_{ab}\frac{\dot{\bar z}_m}{c}-\g_{ab})\hat\de^1_{ba}-ig_c\hat a-i\W\hat\de^1_{bc}+\hat {F}_{ba}+\hat T_{ba},
\ea
\ba
\dot{\hat\de}^1_{bc}=(i\w_{ab}\frac{\dot{\bar z}_m}{c}-\g_{bc})\hat\de^1_{bc}-i\W^*\hat\de^1_{ba}+\hat{F}_{bc}+\hat T_{bc}.
\ea\label{amg-4}
\end{subequations}
where $\hat T_{ba}=i\frac{\w_{ab}\hat F_{th}}{c\nu/Q}\bar\s_{ba}$ and $\hat T_{bc}= i\frac{\w_{ab}\hat F_{th}}{c\nu/Q}\bar{\s}_{bc}.$ The terms $\w_{ab}\dot{\bar z}_m/c,$ $\w_{cb}\dot{\bar z}_m/c,$ $\hat\de^1_{cb},$ and $\hat\de^1_{ab}$ are very small, hence the product terms such as $i\w_{ab}\dot{\bar z}_m\hat\de^1_{bc}/c$ and $i\w_{ab}\dot{\bar z}_m\hat\de^1_{ab}/c$, can be neglected in \eq{amg-4}. By using the Fourier transform function $\mathfrak{F}(x(t))=\frac{1}{\sqrt{2\pi}}\inti x(t)e^{i\w t}\dv t$, \eq{amg-mx} can be written as
\ba
\frac{\dho}{\dho z}\hat\de(z,\w)=i\frac{\w}{c}-i\frac{Ng_c^*}{c}\frac{c}{L_q}\hat\de^1_{ba}(z,\w),\label{amg-13}
\ea
where $\hat{\de}_{ba}^1(z,\w)$ can be obtained by solving \eq{amg-4} in the frequency domain. Solving \eq{amg-13} gives fluctuation in the field at the output of the atomic cell as
\ba
\hat\de(L,\w)=\hat\de_o(\w)e^{\A L}+\int\limits_{0}^{L}e^{-\A(z_1-L)}(\hat F(z_1,\w)+\hat T(z_1,\w))\dv z_1,\label{amg-5}
\ea
where $\hat\de_o(\w)=\hat\de(z=0,\w),$
$$\hat F(z_1,\w)=\frac{Ng_c^*}{c}\frac{c}{L_q}\frac{\frac{-\W\hat F_{bc}(z_1,\w)}{(i\w-\g_{bc})}+i\hat F_{ba}(z_1,\w)}{(i\w-\e_{ba})+\frac{\W^2}{(i\w-\g_{bc})}},$$ $$\hat T(z_1,\w)=\frac{Ng_c^*}{c}\frac{c}{L_q}\frac{\frac{-\W\hat T_{bc}(z_1,\w)}{(i\w-\g_{bc})}+i\hat T_{ba}(z_1,\w)}{(i\w-\e_{ba})+\frac{\W^2}{(i\w-\g_{bc})}}$$
and
\ba
\A(\w)=i\frac{\w}{c}+\ka\frac{1}{(i\w-\e_{ba}+\frac{\W^2}{i\w-\g_{bc}})}. \label{amg-tayn}
\ea
The first term on the RHS of \eq{amg-5} gives shot noise, while the second term gives atomic noise and the third term gives noise due to the mechanical motion of the atom-mechanical oscillator. Fluctuation in \eq{amg-h}, denoted by $\hat\De$, is given as
\ba
\hat\De=\hat I_1-\hat I_2-\bra\hat I_1-\hat I_2\ket=\hat{\de}^\+_1\bar a+\hat\de_1\bar a^*+\hat\de^\+\bar a_1+\hat\de\bar a_1^*.
\ea
Variance of $\hat\De(\w),$ where $\hat\De(\w)=\mathfrak{F}(\hat\De)$, is given as 
\ba
\bra \hat\De^\+(\w)\hat\De(\w')\ket=(V_t+V_a)\de(\w+\w'),\label{amg-var}
\ea
where $V_t$, and $V_a$ are the power spectral densities of noise due to the EIT system and noise due to the mechanical motion of the atom-mechanical oscillator, respectively. In the next two subsections, we estimate the noise due to EIT and mechanical motion of the atom-mechanical oscillator.
\subsection{Noise due to EIT system}\label{amg-ss-1}
By defining $\hat\de_a(z,\w)$ as the fluctuation in the field at the output of the atomic cell due to EIT system, from \eq{amg-5}, we can write
\ba
\hat\de_a(L,\w)=\hat\de_o(\w)e^{\A L}+\int\limits_{0}^{L}e^{-\A(z_1-L)}\hat F(z_1,\w)\dv z_1,\label{amg-26}
\ea
Noise in the EIT system can be estimated by calculating $\bra\hat\De_a^\+(\w)\hat\De_a(\w')\ket,$ where $\hat\De_a(\w)=(\hat\de_a(L,\w)\bar a_1^*+\hat\de_a^\+(L,-\w)\bar a_1).$ Substituting \eq{amg-26} in $\hat\De_a(\w)$, and by using the correlation functions
$$\bra \hat F_{bc}(t_1,z_1)\hat F_{bc}^\+(t_2,z_2) \ket=2\g_{bc}\de(t_1-t_2)\de(z_1-z_2)\frac{L_q}{N},$$
$$\bra \hat F_{ba}(t_1,z_1)\hat F_{ba}^\+(t_2,z_2) \ket=2\g_{ba}\de(t_1-t_2)\de(z_1-z_2)\frac{L_q}{N},$$
$$\bra \hat\de_o(t)\hat\de_o^\+(t')\ket=\bra \hat\de_1(t)\hat\de_1^\+(t')\ket=\de(t-t').$$
we can write
\begin{widetext}
\ba\bs
V_a=|\bar a|^2+\Big(e^{2\A'L}+\ka\frac{1-e^{-2\A'L}}{-2\A'}\frac{\W^2(-2\g_{bc})+|\g_{bc}|^2(-2\e_{ba})}{|(i\w-\e_{ba})(i\w-\g_{bc})+\W^2|^2}\Big)|\bar a_1|^2,
\label{amg-6}
\end{split}\ea
\end{widetext}
where $\A'$ is the real part of $\A$. The first two terms on the RHS of \eq{amg-6} represent the effect of shot noise, while the last term comes from the atomic noise. Because the signal in \eq{amg-10} is oscillating with frequency $\nu$, we are interested in finding the noise component at $\w=2\pi\nu$. However, we assume that $\nu$ is much less than the bandwidth of EIT window, also the signal in \eq{amg-sig} is evaluated under the condition $t_m\ll\nu$. Hence we do a Taylor expansion of \eq{amg-6} and keep the terms only up to the first order of $\nu$. Substituting \eq{amg-tayn} and \eq{amg-8} in \eq{amg-6}, at $\W^2=\ka\g_{bc}L$,  we obtain
\ba
V_1=|\bar a_o|^2(1+e^{-2}),\label{amg-v1}
\ea
where $V_1$ is power spectral density of noise due to EIT system when $\W^2=\ka\g_{bc}L$.
\subsection{Noise due to mechanical oscillator}\label{amg-ss-2}
We define $\hat\de_t$ as the fluctuation in the field, at the output of the atomic cell, due to $\hat F_{th}$. From \eq{amg-5}, we can write
\ba
\hat\de_t(L,\w)=\int\limits_{0}^{L}e^{-\A(z_1-L)}\hat T(z_1,\w))\dv z_1\label{amg-3}
\ea
Substituting $\hat T_{ba}=i\frac{\w_{ab}\hat F_{th}}{c\nu/Q}\bar\s_{ba}$ and $\hat T_{bc}= i\frac{\w_{ab}\hat F_{th}}{c\nu/Q}\bar{\s}_{bc},$ in \eq{amg-3} and by using the relation $\bar \s_{bc}=i\W^*\bar\s_{ba}/(i\w_{cb}\dot{\bar z}/c-\g_{bc})$, we can write that 
\begin{widetext}
\ba
\hat\de_t(L,\w)=\frac{Ng_c}{c}\frac{c}{L_q}\frac{(\frac{\w_{ab}\W^2}{(i\w_{cb}\frac{\dot{\bar z}_m}{c}-\g_{bc})(i\w-\g_{bc})}-\w_{ab})\frac{\hat F_{th}(\w)}{c\nu/Q}\nt{0}{L}e^{-\A(z_1-L)}\bar{\s}_{ba}(z_1)\dv z_1}{(i\w-\e_{ba}+\frac{\W^2}{i\w-\g_{bc}})},\label{amg-9}
\ea
From \eq{amg-sba} and \eq{amg-8}, we can write
\ba\bs
\nt{0}{L}e^{\A(L-z_1)}\bar\s_{ba}(z_1)\dv z_1=\frac{ig_c\A_oe^{\A L}}{\ka}\nt{0}{L}e^{(\A_o-\A)z_1}\bar a_o\dv z_1\approx\frac{ig_c\A_o}{\ka}\bar a_oe^{\A L}L.\label{amg-12}
\end{split}\ea
\end{widetext}
By substituting \eq{amg-12} in \eq{amg-9}, noise due to $\hat F_{th}$ can be estimated by calculating $\bra\hat\De_t^\+(\w)\hat\De_t(\w')\ket,$ where $\hat\De_t(\w)=(\hat\de_t(L,\w)\bar a_1^*+\hat\de_t^\+(L,-\w)\bar a_1).$ 
\begin{widetext}
\ba\bs
\bra\hat\De_t^\+(\w)\hat\De_t(\w)\ket=|\frac{\W^2\A_oe^{\A L}}{(i\w_{cb}\frac{\dot{\bar z}_m}{c}-\g_{bc})(i\w-\g_{bc})}-2\A_oe^{\A L}+\frac{\W^2\A^*_oe^{\A^* L}}{(-i\w_{cb}\frac{\dot{\bar z}_m}{c}-\g_{bc})(i\w-\g_{bc})}-2\A^*_oe^{\A^* L}|^2\\
\times|\frac{\w_{cb}\bar a_1}{(i\w-\e_{ba})+\frac{\W^2}{i\w-\g_{bc}}}|^2\bra\hat F_{th}\hat F_{th}\ket(\frac{Q}{c\nu})^2|\bar a_o|^2=V_t\de(\w+\w'). \label{amg-15}
\end{split}\ea
\end{widetext}
Applying the Taylor expansion on \eq{amg-15} and keeping the terms up to the first order of $\nu$ and $\dot{\bar z}_m$, and using the correlation function $$\bra\hat F_{th}(\w)\hat F_{th}(\w')\ket=\frac{h\nu^2}{Qm}\Big(1+\frac{2}{e^{h\nu/k_BT}-1}\Big)\de(\w+\w'),$$
where $h$ is Planck constant, $m$ is effective mass of atom-mechanical oscillator, $k_B$ is the Boltzmann constant, $T$ is the temperature of the mechanical oscillator along Z-axis, we can simplify $V_t$ as
\ba
V_t\approx 4|\bar a_o|^4\frac{Q\w^2_{ab}h}{c^2\g^2_{bc}m}\Big(1+\frac{2}{e^{h\nu/k_BT}-1}\Big)e^{-2}.\label{amg-vt}
\ea
We used the condition $\W^2=\ka\g_{bc}L$ in \eq{amg-vt}. For
\ba
4|\bar a_o|^2\frac{Q\w^2_{ab}h}{c^2\g^2_{bc}m}(1+\frac{2}{e^{h\nu/k_BT}-1})=1,\label{amg-tc}
\ea
the noise due to the mechanical motion of the atomic cell is in the same order of magnitude as the shot noise. So when \eq{amg-tc} is satisfied, from \eq{amg-vt}, we can write
\ba
V_{2}=|\bar a_o|^2e^{-2},\label{amg-v2}
\ea
where $V_2$ is the power spectral density of noise due to the motion of atomic cell when \eq{amg-tc} is satisfied.
\section{Results and discussion}
The total noise is given by shot noise, atomic noise and noise due to the mechanical motion of the atomic cell. Hence, the total noise can be obtained by from \eq{amg-v1} and \eq{amg-v2} as
\ba
N_o=\sqrt{V_1+V_2}=\sqrt{|\bar a_o|^2}\sqrt{1+2e^{-2}},
\ea
where $N_o$ is the total noise when $\W^2=\ka\g_{bc}L$ and when \eq{amg-tc} is satisfied. The minimum detectable angular velocity $\dot{\tx}_m:=\dot{\tx}N_o/S_o$ is estimated as
\ba
\dot{\tx}_m=\frac{\nu c \g_{bc}}{4Q\dot{\bar y}_o\w_{cb}}\sqrt{\frac{\hbar\w_p}{Pt_m}}\sqrt{1+2e^2}, \label{amg-sen}
\ea
where $P$ is the power of the input probe field corresponding to the condition given in \eq{amg-tc}. Considering the realistic parameters such as $\dot{\bar y}_o=0.01$\,m/s, $c=3\times 10^{8}$\,m/s, $Q=10^7$, $\nu=1$\,Hz, $\w_{ab}/2\pi=9\times10^{14}$\,Hz, $\g_{bc}=10^3$\,Hz, $\hbar=1.054\times 10^{-34}$\,J$\cdot$s, $m=1$\,Kg, $t_m=0.1$\,s, $k_B=1.38\times 10^{-23}$\,J/K, $N=5\times10^{18}$\,m$^{-3}$ we estimate the quantum limit of minimum detectable rotation rate as $\dot{\tx}_m=8.1\times 10^{-19}$\,rad/s at $P=1.6$\,Watt. At $T=300$\,K, the minimum detectable rotation rate is estimated as $\dot{\tx}_m=2.3\times10^{-12}$\,rad/s for $P=2\times10^{-14}$\,Watt. 

From \eq{amg-sen}, we can improve the gyroscope's sensitivity by increasing $\dot{\bar y}_o$. It is also possible to improve the sensitivity by decreasing $\g_{bc}$ or by increasing $Q$, provided that the relation given \eq{amg-tc} is maintained, otherwise the thermal noise from the mechanical oscillator can decrease the sensitivity. One may also wonder about the Fresnel-Fizeau dragging of light by the atomic medium. Especially, when a dispersive system like EIT system is considered~\cite{carusotto,sankar-pra1,safari}. However, according to \eq{amg-tay}, calculation in this manuscript is strictly limited to the linear dispersion case only. Hence the phase velocity of light inside the moving atomic cell is given as
\ba
\frac{c}{\x(\w_p-\dot{\hat z}_m\frac{\w_p}{c})}\approx\frac{c}{\x(\w_p)}+\dot{\hat z}_m\frac{\w_p}{\x(\w_p)^2}\fd{\x}{\w}|_{\w=\w_p}. \label{amg-f}
\ea
When non-linear dispersion is neglected, and for $\x(\w)\approx 1$ (which is the case for EIT at two-photon resonance), the Fresnel-Fizeau dragging is equal to the second term on the RHS of \eq{amg-f}.

Most of the modern navigation grade gyroscopes are based on the Sagnac effect~\cite{post, malykin, malykin-ss}. Sagnac based RLG\cite{chow, faucheux} is limited by the line-width~\cite{cresser,schleich} of the laser field. Also, at very small rotations rates, cross-talk~\cite{cresser, cresser-2} between the co-propagating and counter propagating laser modes reduces the accuracy of these gyroscopes. The sensitivity of FOG~\cite{lefevre,vali} depends on the length of the optical path. Generally, high sensitivity is achieved in FOGs by passing the laser through a long optical fiber~\cite{schiller}. So the sensitivity of FOG is limited because of the scattering and absorption of light inside the optical fiber. Matter wave gyroscope has better sensitivity~\cite{marlan-mwg, campbell, wu, qi} when compared with FOG and ALG because of the short wavelength of matter wave in comparison with optical wave length. All the Sagnac effect based gyroscopes, like RLG, FOG, and matter wave gyroscopes, are dependent on the dimension of the Sagnac loop~\cite{hurst}. Unlike the Sagnac effect based gyroscopes, the atom-mechanical gyroscope do not depend on the dimensions. Quantum spin based rotation sensors have reached ultra-high sensitivity, however, these systems are difficult to commercialize. Conventional two-dimensional mechanical gyroscopes detect rotation by measuring Coriolis force. Radiation pressure force~\cite{caves-rp}, which is exerted by the probe laser, limits the sensitivity of these systems. In the atom-mechanical gyroscope, the probe laser is almost transparent because of EIT and hence the radiation pressure force do not have any significant effect in this system.

\subsection{Doppler broadening effect}
Even when there is no rotation, at $\dot\tx=0$, atoms inside the atomic cell are moving randomly in all directions with finite velocity because of thermal energy. This thermal motion, which is different from the Doppler effect arising due to rotation, leads to Doppler broadening effect. In this section we estimate the effect of thermal Doppler broadening. In presence of thermal motion, the $\bar\s_{ba}$ is given as
\ba
\bar\s^1_{baD}=\frac{ig_c\bar a(i\w_{cb}\frac{\dot{\bar z}_m}{c}+i\De_3-\g_{bc})}{(i\w_{ab}\frac{\dot{\bar z}_m}{c}+i\De_1-\g_{ab})(i\w_{cb}\frac{\dot{\bar z}_m}{c}+i\De_3-\g_{cb})+|\W|^2},\label{amg-dop}
\ea
where $\De_1=\w_{ab}v_z/c$, $\De_3=(\w_{ab}v_z-(\w_{ab}-\w_{cb})v_y)/c$ represent the frequency detuning due to thermal Doppler effect. $v_z$ and $v_y$ are the velocity components of an atom along probe laser and drive laser, respectively, $\bar\s^1_{baD}$ represents $\bar\s_{ba}$ in presence of thermal Doppler effect. Assuming that the cold atoms are at a temperature $T_a=10^{-6}$\ K, and have mass $m_a=1.4\times10^{-25}$\ Kg, the most probable velocity of an atom is given as $\sqrt{2k_BT_a/m_a}\approx 10^{-2}$\ m/s. We estimate that $\De_1\approx10^5$\ Hz and $\De_3\approx10^{-1}$\,Hz for $\w_{cb}=10^{-6}\w_{ab}$. Hence by using the approximation $\W^2\gg\De_1\g_{bc}+\De_3\g_{ab}\gg\De_1\De_3$, \eq{amg-dop} can be simplified as 
\ba
\bar\s_{baD}\approx ig_c\bar a(\frac{(i\De_3-\g_{bc})}{|\W|^2}+i\frac{\w_{cb}\dot{\bar z}_m/c}{|\W|^2}).\label{amg-sbad}
\ea
Using the Maxwellian velocity distribution, Doppler broadening effect is estimated as
\ba
\frac{m}{2\pi k_BT}\inti\inti e^{-m(v_x^2+v_y^2)/2k_BT}\bar\s_{baD}\dv v_y\dv v_z=\bar\s_{ba}. 
\ea
Hence thermal Doppler effect is zero for the cold atoms. Note that the thermal motion of atoms is randomly directed in all directions and this leads to insignificant effect when integrated over all the possible directions. On the other hand, motion of the atom-mechanical oscillator due to Coriolis effect is non-random. Hence the velocity of the atom-mechanical oscillator due to the Coriolis force gives finite signal even in presence of thermal Doppler effect.
\subsection{Effect of cooling lasers on atom-mechanical oscillator}
Generally, cold atoms are prepared~\cite{hansch, wineland} by reducing the atomic thermal kinetic energy by using a set of laser fields in all the three dimensions. So it is important to consider the effect of these cooling lasers on the velocity of atom-mechanical oscillator. The atom-mechanical  is driven with velocity $\dot{\bar y}$ along Y-axis. We can fix the cooling laser set-up on the driving platform so the cooling lasers and the atom-mechanical are not moving with respect to each other along Y-axis.

Assuming that the cold atoms are at a temperature $T_a=10^{-6}$\ K and have mass $m_a=1.4\times10^{-25}$\ Kg, the most probable velocity of an atom is about $ 10^{-2}$\,m/s. The maximum velocity of the atomic cloud due to the action of Coriolis force is 
\ba
\nt{0}{t_m}2\dot\tx\dot{\bar y}\dv t=\nt{0}{t_m}2\dot\tx\dot{\bar y}_o\cos(\w_mt)\dv t\approx2\dot{\bar y}_o\dot{\tx}t_m.
\ea
Substituting $\dot{\bar y}_o=10^{-2}$\,m/s, $t_m=0.1$\,s, and $\dot{\tx}=10^{-11}$\,rad/sec, we note that the maximum velocity of atom-mechanical oscillator along Z-axis is about $10^{-14}$\,m/s. Hence kinetic energy of atoms due to the Coriolis force is much smaller than the thermal kinetic energy of atoms and therefore the cooling lasers can not hinder the motion of the atom-mechanical oscillator due to the Coriolis force. 
\subsection{Effect of fluctuations in mechanical driving along Y-axis}
In \eq{amg-1}, we assumed that the mechanical drive along Y-axis is classical and it has no fluctuations. Such an approximation is accurate with in the linear regime for the following reason. Suppose that $\de_{\dot y}$ is the fluctuation in the velocity of driving platform along Y-axis, then \eq{amg-vel} becomes
\ba
\dot{\hat z}_m=\frac{2Q(\dot{\bar y}+\de_{\dot y})\dot\tx}{\nu}\cos(2\pi\nu t)+\frac{Q}{\nu}\hat F_{th}.\label{amg-vela}
\ea
Both $\dot\tx$ and $\de_{\dot y}$ are small and hence their product term can be neglected. Hence \eq{amg-vela} reduces to \eq{amg-vel} even when fluctuations along Y-axis are considered. Moreover, we are interested in measuring classical Coriolis force and hence it is reasonable to drop any fluctuation in the driving platform's velocity. Similarly we can also show that the thermal fluctuations in driving platform do not effect the rotation detection sensitivity. Given the recent progress in cooling\cite{arcizet,marquardt, bhattacharya,schliesser-cooling} mechanical oscillators to their ground state, quantum limited sensitivity of $8\times10^{-19}$rad/s can be realized. Hence the method described in this work can be used to realize gyroscopes with very high sensitivities.

\section{Conclusion}
Application of hybrid atom-mechanical system for absolute rotation detection is studied. Quantum noise limited sensitivity of the atom-mechanical gyroscope is estimated as $8\times 10^{-19}$\,rad/s. For optimal performance of the gyroscope, the condition under which thermal noise from the mechanical oscillator can be set to the level of shot noise is derived. For the cold-atom set-up, assuming that the atoms are at $T_a=10^{-6}$\,K, the rotation detection sensitivity is estimated as $7\times10^{-12}$\,rad/s when the mechanical oscillator is at room temperature ($T=300$\,K).
\section{Acknowledgments}
This work is supported by the National Key R\&D Program of China grant 2016YFA0301200 and the National Basic Research Program of China (under Grant No. 2014CB921403). It is also supported by Science Challenge Project (under Grant No. TZ2017003) and the National Natural Science Foundation of China (under Grants No. 11774024, No. 11534002, and No. U1530401)."
\appendix
\section{Many atom formalism}\label{amg-a1}
From \eq{amg-ham}, the equation for motion for the field mode $\hat c_r$ is given as
\ba
\dot{\hat c}_r=-i\w_r\hat c_r-ig^*\sum_{s}^{}\s^{ls}_{ba}e^{-i(k_r-k_p)z_l},\label{amg-20}
\ea
By substituting $\w_r=\w_p+\frac{2r\pi c}{L_q}$ in \eq{amg-20} and by taking summation on all the field modes, we can rewrite \eq{amg-20} as
\begin{widetext}
\begin{equation}\bs
\implies&\pd{}{t}\um{r}{}\hat c_re^{i(k_r-k_p)z_l}=-i\um{r}{}(\w_p+\frac{2r\pi c}{L_q})\hat c_re^{i(k_r-k_p)z_l}-i\um{r,s}{}g_r^*\s^{ls}_{ba},\\
\implies&\pd{}{t}\um{r}{}\hat c_re^{i(k_r-k_p)z_l}=-i\um{r}{}(\w_p+c\pd{}{z_l})\hat c_re^{i(k_r-k_p)z_l}-i\um{r,s}{}g_r^*\s^{ls}_{ba}.\label{amg-24}
\end{split}\end{equation}
\end{widetext}
By defining a new operator $\hat c_l(z_l,t)$ as $$\hat c_l(z_l,t)=\um{r}{}\hat c_re^{i(k_r-k_p)z_l},$$ \eq{amg-24} can be rewritten as
\begin{subequations}
\ba\bs
\pd{}{t}\hat c_l(z_l,t)=-i\w_p\hat c_l-c\pd{}{z_l}\hat c_l(z_l,t)-i\um{r,s}{}g_r^*\s^{ls}_{ba}.\label{amg-23}
\end{split}\ea
Equations of motion for the atomic operators are
\ba\bs
\dot{\hat\s}^{ls}_{ba}=(-i\w_{ab}-\g_{ab})\hat\s^{ls}_{ba}+ig_r(\hat\s^{ls}_{aa}-\hat\s^{ls}_{bb})\hat c_l -i\td\W\hat\s^{ls}_{bc}+\hat F^{ls}_{ba},
\end{split}\ea
\ba\bs
\dot {\hat\s}^{ls}_{bc}=(-i\w_{cb}-\g_{bc})\hat\s^{ls}_{bc}+ig_r\hat\s^{ls}_{ac}\hat c_l-i\td\W^*\hat\s^{ls}_{ba}+\hat F^{ls}_{bc},
\end{split}\ea
\ba\bs
\dot{\hat\s}^{ls}_{ac}=(i\w_{ac}-\g_{ac})\hat\s^{ls}_{ac}+i(\hat\s^{ls}_{cc}-\hat\s^{ls}_{aa})\td\W^*+ig_r^*\hat\s^{ls}_{bc}\hat c_l^{\+}+\hat F^{ls}_{ac},
\end{split}\ea
\end{subequations}
$\hat F^{ls}_{uv}$ is the single atom noise operator.Assuming that the band-width of quasi-monochromatic probe field is much less than its mean frequency $\w_p$, we can write $\sum_{r}{}g_r=(2R+1)g$, with $g=\sqrt{\hbar\w_p/\E_oAL_q}$. Hence \eq{amg-23} and \eq{amg-24} can be written as
\begin{widetext}
\begin{subequations}
\ba
\pd{}{t}\hat c_l(z_l,t)=-i\w_p\hat c_l-c\pd{}{z_l}\hat c_l(z_l,t)-i(2R+1)\um{s}{}g^*\s^{ls}_{ba},
\ea
\ba\bs
(2R+1)\um{s}{}\dot{\hat\s}^{ls}_{ba}=(-i\w_{ab}-\g_{ab})(2R+1)\um{s}{}\hat\s^{ls}_{ba}+ig\um{s}{}(2R+1)(\hat\s^{ls}_{aa}-\hat\s^{ls}_{bb})\hat c_l-\um{s}{}(2R+1)i\td\W\hat\s^{ls}_{bc}+\um{s}{}(2R+1)\hat F^{ls}_{ba},
\end{split}\ea
\ba\bs
\um{s}{}(2R+1)\dot {\hat\s}^{ls}_{bc}=(-i\w_{cb}-\g_{bc})\um{s}{}(2R+1)\hat\s^{ls}_{bc}+ig\um{s}{}(2R+1)\hat\s^{ls}_{ac}\hat c_l-\um{s}{}(2R+1)i\td\W^*\hat\s^{ls}_{ba}+\um{s}{}(2R+1)\hat F^{ls}_{bc},
\end{split}\ea
\ba\bs
\um{s}{}(2R+1)\dot{\hat\s}^{ls}_{ac}=(i\w_{ac}-\g_{ac})\um{s}{}(2R+1)\hat\s^{ls}_{ac}+\um{s}{}(2R+1)i(\hat\s^{ls}_{cc}-\hat\s^{ls}_{aa})\td\W^*+ig^*\um{s}{}(2R+1)\hat\s^{ls}_{bc}\hat c_l^{\+}+\um{s}{}(2R+1)\hat F^{ls}_{ac},
\end{split}\ea
\label{amg-18}
\end{subequations}
\end{widetext}
In the limit $R\to\infty$ or $\De z:=z_{l+1}-z_l\to 0$, we go from discreet to continuous formulation by using the following transformations
\begin{subequations}
\ba z_l=\frac{lL_q}{2P+1}\to z,\ea
\ba\um{\w_r}{}\to\frac{1}{\w_{r+1}-\w_r}\int\dv\w,\ea
\ba\hat c_r\to\sqrt{\w_{r+1}-\w_r}\hat c(\w)=\sqrt{\De\w}\hat c(\w)\ea
\ba\frac{2R+1}{N}\um{s}{}\hat\s_{ij}^{ls}\to\frac{2R+1}{N}\um{s}{}\hat\s_{ij}^{ls}|_{z_l\to z}:=\hat\s_{ij}^o,\ea
\ba\frac{2R+1}{N}\um{s}{}\hat F_{ij}^{ls}\to\frac{2R+1}{N}\um{s}{}\hat F_{ij}^{ls}|_{z_l\to z}:=\hat F^o_{ij}\ea
\ba\frac{\de_{z,z'}}{z_{p+1}-z_p}=\frac{\de_{z,z'}}{\De z}\to\de(z-z'),\ea
\ba\frac{\de_{\w,\w'}}{\w_{r+1}-\w_r}=\frac{\de_{\w,\w'}}{\De\w}\to\de(\w-\w').\ea \label{amg-dc1}
\end{subequations}
Using \eq{amg-dc1}, we can write
\ba\bs
&\um{r}{}g_r\hat c_re^{i(k_r-k_p)z_l}=\um{r}{}g_r\hat{\td c}_re^{i(k_r-k_p)z_l}e^{-i\w_rt}\\
&\to\sqrt{\frac{\hbar\w_p}{\E_oAc}}\frac{1}{\sqrt{2\pi}}\nt{0}{\infty}\dv\w \hat{\td c}(\w)e^{i(k-k_p)z}e^{-i\w t}\\
&:=\sqrt{\frac{\hbar\w_p}{\E_oAc}}\hat ae^{-i\w_pt}, \label{amg-22}
\end{split}\ea
We further defined a new operator $\hat a$ which is normalized such that $\bra\hat  a^\+\hat a\ket$ represents the photons per unit time. By using \eq{amg-22} and the relation $\um{r}{}g_r=(2R+1)g$, \eq{amg-18} can be transformed from discreet to continuous formalism. In the continuous notation, equations of motion are given as
\begin{subequations}
\ba
\Big(\frac{\dho}{\dho t}+c\frac{\dho}{\dho z}\Big)\hat a=-iN\sqrt\frac{c}{L_q}g^*\hat\s_{ba}^oe^{i\w_pt},
\ea
\ba
\dot{\hat\s}^o_{ba}=(-i\w_{ab}-\g_{ab})\hat\s^o_{ba}+ig_c(\hat\s^o_{aa}-\hat\s^o_{bb})\hat ae^{-i\w_pt}-i\td\W\hat\s^o_{bc}+\hat F^o_{ba},
\ea
\ba
\dot {\hat\s}^o_{bc}=(-i\w_{cb}-\g_{bc})\hat\s^o_{bc}+ig_c\hat\s^o_{ac}\hat ae^{-i\w_pt}-i\td\W^*\hat\s^o_{ba}+\hat F^o_{bc},
\ea
\ba
\dot{\hat\s}^o_{ac}=(i\w_{ac}-\g_{ac})\hat\s^o_{ac}+i(\hat\s^o_{cc}-\hat\s^o_{aa})\td\W^*+ig_c^*\hat\s^o_{bc}\hat a^{\+}e^{i\w_pt}+\hat F^o_{ac},
\ea
\label{amg-21}
\end{subequations}
After using rotating wave approximation by writing $\hat \s^o_{ba}=\hat \s_{ba}e^{-i\w_pt}$ and $\hat \s^o_{ca}=\hat\s_{ca}e^{-i\w_dt}$, we can write
\begin{subequations}
\ba
\Big(\frac{\dho}{\dho t}+c\frac{\dho}{\dho z}\Big)\hat a=-iNg^*\sqrt\frac{c}{L_q}\hat\s_{ba},
\ea
\ba
\dot{\hat\s}_{ba}=[-i(\w_{ab}-\w_p)-\g_{ab}]\hat\s_{ba}+ig_c(\hat\s_{aa}-\hat\s_{bb})\hat a-i\W\hat\s_{bc}+\hat F_{ba},
\ea
\ba
\dot {\hat\s}_{bc}=[-i(\w_{cb}-(\w_p-\w_d))-\g_{bc}]\hat\s_{bc}+ig_c\hat\s_{ac}\hat a-i\W\hat\s_{ba}+\hat F_{bc},
\ea
\ba
\dot{\hat\s}_{ac}=[i(\w_{ac}-\w_d)-\g_{ac}]\hat\s_{ac}+i(\hat\s_{cc}-\hat\s_{aa})\W+ig_c^*\hat\s_{bc}\hat a^{\+}+\hat F_{ac}.
\ea
\label{amg-17}
\end{subequations}
\bibliography{references}

\begin{thebibliography}{62}%
\makeatletter
\providecommand \@ifxundefined [1]{%
 \@ifx{#1\undefined}
}%
\providecommand \@ifnum [1]{%
 \ifnum #1\expandafter \@firstoftwo
 \else \expandafter \@secondoftwo
 \fi
}%
\providecommand \@ifx [1]{%
 \ifx #1\expandafter \@firstoftwo
 \else \expandafter \@secondoftwo
 \fi
}%
\providecommand \natexlab [1]{#1}%
\providecommand \enquote  [1]{``#1''}%
\providecommand \bibnamefont  [1]{#1}%
\providecommand \bibfnamefont [1]{#1}%
\providecommand \citenamefont [1]{#1}%
\providecommand \href@noop [0]{\@secondoftwo}%
\providecommand \href [0]{\begingroup \@sanitize@url \@href}%
\providecommand \@href[1]{\@@startlink{#1}\@@href}%
\providecommand \@@href[1]{\endgroup#1\@@endlink}%
\providecommand \@sanitize@url [0]{\catcode `\\12\catcode `\$12\catcode
  `\&12\catcode `\#12\catcode `\^12\catcode `\_12\catcode `\%12\relax}%
\providecommand \@@startlink[1]{}%
\providecommand \@@endlink[0]{}%
\providecommand \url  [0]{\begingroup\@sanitize@url \@url }%
\providecommand \@url [1]{\endgroup\@href {#1}{\urlprefix }}%
\providecommand \urlprefix  [0]{URL }%
\providecommand \Eprint [0]{\href }%
\providecommand \doibase [0]{http://dx.doi.org/}%
\providecommand \selectlanguage [0]{\@gobble}%
\providecommand \bibinfo  [0]{\@secondoftwo}%
\providecommand \bibfield  [0]{\@secondoftwo}%
\providecommand \translation [1]{[#1]}%
\providecommand \BibitemOpen [0]{}%
\providecommand \bibitemStop [0]{}%
\providecommand \bibitemNoStop [0]{.\EOS\space}%
\providecommand \EOS [0]{\spacefactor3000\relax}%
\providecommand \BibitemShut  [1]{\csname bibitem#1\endcsname}%
\let\auto@bib@innerbib\@empty
%</preamble>
\bibitem [{\citenamefont {Everitt}\ \emph {et~al.}(2011)\citenamefont
  {Everitt}, \citenamefont {DeBra}, \citenamefont {Parkinson}, \citenamefont
  {Turneaure}, \citenamefont {Conklin}, \citenamefont {Heifetz}, \citenamefont
  {Keiser}, \citenamefont {Silbergleit}, \citenamefont {Holmes}, \citenamefont
  {Kolodziejczak}, \citenamefont {Al-Meshari}, \citenamefont {Mester},
  \citenamefont {Muhlfelder}, \citenamefont {Solomonik}, \citenamefont {Stahl},
  \citenamefont {Worden}, \citenamefont {Bencze}, \citenamefont {Buchman},
  \citenamefont {Clarke}, \citenamefont {Al-Jadaan}, \citenamefont
  {Al-Jibreen}, \citenamefont {Li}, \citenamefont {Lipa}, \citenamefont
  {Lockhart}, \citenamefont {Al-Suwaidan}, \citenamefont {Taber},\ and\
  \citenamefont {Wang}}]{everitt}%
  \BibitemOpen
  \bibfield  {author} {\bibinfo {author} {\bibfnamefont {C.~W.~F.}\
  \bibnamefont {Everitt}}, \bibinfo {author} {\bibfnamefont {D.~B.}\
  \bibnamefont {DeBra}}, \bibinfo {author} {\bibfnamefont {B.~W.}\ \bibnamefont
  {Parkinson}}, \bibinfo {author} {\bibfnamefont {J.~P.}\ \bibnamefont
  {Turneaure}}, \bibinfo {author} {\bibfnamefont {J.~W.}\ \bibnamefont
  {Conklin}}, \bibinfo {author} {\bibfnamefont {M.~I.}\ \bibnamefont
  {Heifetz}}, \bibinfo {author} {\bibfnamefont {G.~M.}\ \bibnamefont {Keiser}},
  \bibinfo {author} {\bibfnamefont {A.~S.}\ \bibnamefont {Silbergleit}},
  \bibinfo {author} {\bibfnamefont {T.}~\bibnamefont {Holmes}}, \bibinfo
  {author} {\bibfnamefont {J.}~\bibnamefont {Kolodziejczak}}, \bibinfo {author}
  {\bibfnamefont {M.}~\bibnamefont {Al-Meshari}}, \bibinfo {author}
  {\bibfnamefont {J.~C.}\ \bibnamefont {Mester}}, \bibinfo {author}
  {\bibfnamefont {B.}~\bibnamefont {Muhlfelder}}, \bibinfo {author}
  {\bibfnamefont {V.~G.}\ \bibnamefont {Solomonik}}, \bibinfo {author}
  {\bibfnamefont {K.}~\bibnamefont {Stahl}}, \bibinfo {author} {\bibfnamefont
  {P.~W.}\ \bibnamefont {Worden}}, \bibinfo {author} {\bibfnamefont
  {W.}~\bibnamefont {Bencze}}, \bibinfo {author} {\bibfnamefont
  {S.}~\bibnamefont {Buchman}}, \bibinfo {author} {\bibfnamefont
  {B.}~\bibnamefont {Clarke}}, \bibinfo {author} {\bibfnamefont
  {A.}~\bibnamefont {Al-Jadaan}}, \bibinfo {author} {\bibfnamefont
  {H.}~\bibnamefont {Al-Jibreen}}, \bibinfo {author} {\bibfnamefont
  {J.}~\bibnamefont {Li}}, \bibinfo {author} {\bibfnamefont {J.~A.}\
  \bibnamefont {Lipa}}, \bibinfo {author} {\bibfnamefont {J.~M.}\ \bibnamefont
  {Lockhart}}, \bibinfo {author} {\bibfnamefont {B.}~\bibnamefont
  {Al-Suwaidan}}, \bibinfo {author} {\bibfnamefont {M.}~\bibnamefont {Taber}},
  \ and\ \bibinfo {author} {\bibfnamefont {S.}~\bibnamefont {Wang}},\ }\href
  {\doibase 10.1103/PhysRevLett.106.221101} {\bibfield  {journal} {\bibinfo
  {journal} {Phys. Rev. Lett.}\ }\textbf {\bibinfo {volume} {106}},\ \bibinfo
  {pages} {221101} (\bibinfo {year} {2011})}\BibitemShut {NoStop}%
\bibitem [{\citenamefont {Stedman}(1997)}]{stedman}%
  \BibitemOpen
  \bibfield  {author} {\bibinfo {author} {\bibfnamefont {G.~E.}\ \bibnamefont
  {Stedman}},\ }\href {http://stacks.iop.org/0034-4885/60/i=6/a=001} {\bibfield
   {journal} {\bibinfo  {journal} {Reports on Progress in Physics}\ }\textbf
  {\bibinfo {volume} {60}},\ \bibinfo {pages} {615} (\bibinfo {year}
  {1997})}\BibitemShut {NoStop}%
\bibitem [{\citenamefont {Smith}(1987)}]{smith}%
  \BibitemOpen
  \bibfield  {author} {\bibinfo {author} {\bibfnamefont {S.}~\bibnamefont
  {Smith}},\ }\href {http://stacks.iop.org/0305-4624/18/i=4/a=I03} {\bibfield
  {journal} {\bibinfo  {journal} {Physics in Technology}\ }\textbf {\bibinfo
  {volume} {18}},\ \bibinfo {pages} {165} (\bibinfo {year} {1987})}\BibitemShut
  {NoStop}%
\bibitem [{\citenamefont {Lee}\ and\ \citenamefont {Han}(2009)}]{lee-is}%
  \BibitemOpen
  \bibfield  {author} {\bibinfo {author} {\bibfnamefont {D.-C.}\ \bibnamefont
  {Lee}}\ and\ \bibinfo {author} {\bibfnamefont {C.-S.}\ \bibnamefont {Han}},\
  }\href {\doibase 10.1243/09544062JMES1233} {\bibfield  {journal} {\bibinfo
  {journal} {Proceedings of the Institution of Mechanical Engineers, Part C:
  Journal of Mechanical Engineering Science}\ }\textbf {\bibinfo {volume}
  {223}},\ \bibinfo {pages} {1687} (\bibinfo {year} {2009})},\ \Eprint
  {http://arxiv.org/abs/http://pic.sagepub.com/content/223/7/1687.full.pdf+html}
  {http://pic.sagepub.com/content/223/7/1687.full.pdf+html} \BibitemShut
  {NoStop}%
\bibitem [{\citenamefont {Ezekiel}(1974)}]{ezekiel-alg}%
  \BibitemOpen
  \bibfield  {author} {\bibinfo {author} {\bibfnamefont {S.}~\bibnamefont
  {Ezekiel}},\ }\href {\doibase 10.1117/12.7978725} {\bibfield  {journal}
  {\bibinfo  {journal} {Optical Engineering}\ }\textbf {\bibinfo {volume}
  {13}},\ \bibinfo {pages} {136217} (\bibinfo {year} {1974})}\BibitemShut
  {NoStop}%
\bibitem [{\citenamefont {Arditty}\ and\ \citenamefont
  {Lef\`{e}vre}(1981)}]{arditty}%
  \BibitemOpen
  \bibfield  {author} {\bibinfo {author} {\bibfnamefont {H.~J.}\ \bibnamefont
  {Arditty}}\ and\ \bibinfo {author} {\bibfnamefont {H.~C.}\ \bibnamefont
  {Lef\`{e}vre}},\ }\href {\doibase 10.1364/OL.6.000401} {\bibfield  {journal}
  {\bibinfo  {journal} {Opt. Lett.}\ }\textbf {\bibinfo {volume} {6}},\
  \bibinfo {pages} {401} (\bibinfo {year} {1981})}\BibitemShut {NoStop}%
\bibitem [{\citenamefont {Chow}\ \emph {et~al.}(1985)\citenamefont {Chow},
  \citenamefont {Gea-Banacloche}, \citenamefont {Pedrotti}, \citenamefont
  {Sanders}, \citenamefont {Schleich},\ and\ \citenamefont {Scully}}]{chow}%
  \BibitemOpen
  \bibfield  {author} {\bibinfo {author} {\bibfnamefont {W.~W.}\ \bibnamefont
  {Chow}}, \bibinfo {author} {\bibfnamefont {J.}~\bibnamefont
  {Gea-Banacloche}}, \bibinfo {author} {\bibfnamefont {L.~M.}\ \bibnamefont
  {Pedrotti}}, \bibinfo {author} {\bibfnamefont {V.~E.}\ \bibnamefont
  {Sanders}}, \bibinfo {author} {\bibfnamefont {W.}~\bibnamefont {Schleich}}, \
  and\ \bibinfo {author} {\bibfnamefont {M.~O.}\ \bibnamefont {Scully}},\
  }\href {\doibase 10.1103/RevModPhys.57.61} {\bibfield  {journal} {\bibinfo
  {journal} {Rev. Mod. Phys.}\ }\textbf {\bibinfo {volume} {57}},\ \bibinfo
  {pages} {61} (\bibinfo {year} {1985})}\BibitemShut {NoStop}%
\bibitem [{\citenamefont {Berg}\ \emph {et~al.}(2015)\citenamefont {Berg},
  \citenamefont {Abend}, \citenamefont {Tackmann}, \citenamefont {Schubert},
  \citenamefont {Giese}, \citenamefont {Schleich}, \citenamefont {Narducci},
  \citenamefont {Ertmer},\ and\ \citenamefont {Rasel}}]{berg}%
  \BibitemOpen
  \bibfield  {author} {\bibinfo {author} {\bibfnamefont {P.}~\bibnamefont
  {Berg}}, \bibinfo {author} {\bibfnamefont {S.}~\bibnamefont {Abend}},
  \bibinfo {author} {\bibfnamefont {G.}~\bibnamefont {Tackmann}}, \bibinfo
  {author} {\bibfnamefont {C.}~\bibnamefont {Schubert}}, \bibinfo {author}
  {\bibfnamefont {E.}~\bibnamefont {Giese}}, \bibinfo {author} {\bibfnamefont
  {W.~P.}\ \bibnamefont {Schleich}}, \bibinfo {author} {\bibfnamefont {F.~A.}\
  \bibnamefont {Narducci}}, \bibinfo {author} {\bibfnamefont {W.}~\bibnamefont
  {Ertmer}}, \ and\ \bibinfo {author} {\bibfnamefont {E.~M.}\ \bibnamefont
  {Rasel}},\ }\href {\doibase 10.1103/PhysRevLett.114.063002} {\bibfield
  {journal} {\bibinfo  {journal} {Phys. Rev. Lett.}\ }\textbf {\bibinfo
  {volume} {114}},\ \bibinfo {pages} {063002} (\bibinfo {year}
  {2015})}\BibitemShut {NoStop}%
\bibitem [{\citenamefont {Gustavson}\ \emph {et~al.}(1997)\citenamefont
  {Gustavson}, \citenamefont {Bouyer},\ and\ \citenamefont
  {Kasevich}}]{gustavson}%
  \BibitemOpen
  \bibfield  {author} {\bibinfo {author} {\bibfnamefont {T.~L.}\ \bibnamefont
  {Gustavson}}, \bibinfo {author} {\bibfnamefont {P.}~\bibnamefont {Bouyer}}, \
  and\ \bibinfo {author} {\bibfnamefont {M.~A.}\ \bibnamefont {Kasevich}},\
  }\href {\doibase 10.1103/PhysRevLett.78.2046} {\bibfield  {journal} {\bibinfo
   {journal} {Phys. Rev. Lett.}\ }\textbf {\bibinfo {volume} {78}},\ \bibinfo
  {pages} {2046} (\bibinfo {year} {1997})}\BibitemShut {NoStop}%
\bibitem [{\citenamefont {Malykin}(2000)}]{malykin}%
  \BibitemOpen
  \bibfield  {author} {\bibinfo {author} {\bibfnamefont {G.~B.}\ \bibnamefont
  {Malykin}},\ }\href {http://stacks.iop.org/1063-7869/43/i=12/a=A03}
  {\bibfield  {journal} {\bibinfo  {journal} {Physics-Uspekhi}\ }\textbf
  {\bibinfo {volume} {43}},\ \bibinfo {pages} {1229} (\bibinfo {year}
  {2000})}\BibitemShut {NoStop}%
\bibitem [{\citenamefont {Kajari}\ \emph {et~al.}(2006)\citenamefont {Kajari},
  \citenamefont {Walser}, \citenamefont {Schleich},\ and\ \citenamefont
  {Delgado}}]{kajari}%
  \BibitemOpen
  \bibfield  {author} {\bibinfo {author} {\bibfnamefont {E.}~\bibnamefont
  {Kajari}}, \bibinfo {author} {\bibfnamefont {R.}~\bibnamefont {Walser}},
  \bibinfo {author} {\bibfnamefont {W.~P.}\ \bibnamefont {Schleich}}, \ and\
  \bibinfo {author} {\bibfnamefont {A.}~\bibnamefont {Delgado}},\ }in\ \href
  {http://www.osapublishing.org/abstract.cfm?URI=LS-2006-JWD48} {\emph
  {\bibinfo {booktitle} {Frontiers in Optics}}}\ (\bibinfo  {publisher}
  {Optical Society of America},\ \bibinfo {year} {2006})\ p.\ \bibinfo {pages}
  {JWD48}\BibitemShut {NoStop}%
\bibitem [{\citenamefont {Davuluri}\ and\ \citenamefont
  {Li}(2016)}]{sankar-njp}%
  \BibitemOpen
  \bibfield  {author} {\bibinfo {author} {\bibfnamefont {S.}~\bibnamefont
  {Davuluri}}\ and\ \bibinfo {author} {\bibfnamefont {Y.}~\bibnamefont {Li}},\
  }\href {http://stacks.iop.org/1367-2630/18/i=10/a=103047} {\bibfield
  {journal} {\bibinfo  {journal} {New Journal of Physics}\ }\textbf {\bibinfo
  {volume} {18}},\ \bibinfo {pages} {103047} (\bibinfo {year}
  {2016})}\BibitemShut {NoStop}%
\bibitem [{\citenamefont {Davuluri}\ \emph {et~al.}(2017)\citenamefont
  {Davuluri}, \citenamefont {Li},\ and\ \citenamefont {Li}}]{sankar-njp2}%
  \BibitemOpen
  \bibfield  {author} {\bibinfo {author} {\bibfnamefont {S.}~\bibnamefont
  {Davuluri}}, \bibinfo {author} {\bibfnamefont {K.}~\bibnamefont {Li}}, \ and\
  \bibinfo {author} {\bibfnamefont {Y.}~\bibnamefont {Li}},\ }\href
  {http://stacks.iop.org/1367-2630/19/i=11/a=113004} {\bibfield  {journal}
  {\bibinfo  {journal} {New Journal of Physics}\ }\textbf {\bibinfo {volume}
  {19}},\ \bibinfo {pages} {113004} (\bibinfo {year} {2017})}\BibitemShut
  {NoStop}%
\bibitem [{\citenamefont {Kai}\ \emph {et~al.}()\citenamefont {Kai},
  \citenamefont {Sankar},\ and\ \citenamefont {Yong}}]{sankar-sc}%
  \BibitemOpen
  \bibfield  {author} {\bibinfo {author} {\bibfnamefont {L.}~\bibnamefont
  {Kai}}, \bibinfo {author} {\bibfnamefont {D.}~\bibnamefont {Sankar}}, \ and\
  \bibinfo {author} {\bibfnamefont {L.}~\bibnamefont {Yong}},\ }\href {\doibase
  https://doi.org/10.1007/s11433-018-9189-6} {\bibfield  {journal} {\bibinfo
  {journal} {SCIENCE CHINA Physics, Mechanics \& Astronomy}\ }\textbf {\bibinfo
  {volume} {(accepted)}},\
  https://doi.org/10.1007/s11433-018-9189-6}\BibitemShut {NoStop}%
\bibitem [{\citenamefont {Hau}\ \emph {et~al.}(1999)\citenamefont {Hau},
  \citenamefont {Harris}, \citenamefont {Dutton},\ and\ \citenamefont
  {Behroozi}}]{hau}%
  \BibitemOpen
  \bibfield  {author} {\bibinfo {author} {\bibfnamefont {L.~V.}\ \bibnamefont
  {Hau}}, \bibinfo {author} {\bibfnamefont {S.~E.}\ \bibnamefont {Harris}},
  \bibinfo {author} {\bibfnamefont {Z.}~\bibnamefont {Dutton}}, \ and\ \bibinfo
  {author} {\bibfnamefont {C.~H.}\ \bibnamefont {Behroozi}},\ }\href
  {http://dx.doi.org/10.1038/17561} {\bibfield  {journal} {\bibinfo  {journal}
  {Nature}\ }\textbf {\bibinfo {volume} {397}},\ \bibinfo {pages} {594 EP }
  (\bibinfo {year} {1999})}\BibitemShut {NoStop}%
\bibitem [{\citenamefont {Novikova}\ \emph {et~al.}(2012)\citenamefont
  {Novikova}, \citenamefont {Walsworth},\ and\ \citenamefont
  {Xiao}}]{novikova}%
  \BibitemOpen
  \bibfield  {author} {\bibinfo {author} {\bibfnamefont {I.}~\bibnamefont
  {Novikova}}, \bibinfo {author} {\bibfnamefont {R.}~\bibnamefont {Walsworth}},
  \ and\ \bibinfo {author} {\bibfnamefont {Y.}~\bibnamefont {Xiao}},\ }\href
  {\doibase 10.1002/lpor.201100021} {\bibfield  {journal} {\bibinfo  {journal}
  {Laser \& Photonics Reviews}\ }\textbf {\bibinfo {volume} {6}},\ \bibinfo
  {pages} {333} (\bibinfo {year} {2012})}\BibitemShut {NoStop}%
\bibitem [{\citenamefont {Sautenkov}\ \emph {et~al.}(2010)\citenamefont
  {Sautenkov}, \citenamefont {Li}, \citenamefont {Rostovtsev},\ and\
  \citenamefont {Scully}}]{sautenkov}%
  \BibitemOpen
  \bibfield  {author} {\bibinfo {author} {\bibfnamefont {V.~A.}\ \bibnamefont
  {Sautenkov}}, \bibinfo {author} {\bibfnamefont {H.}~\bibnamefont {Li}},
  \bibinfo {author} {\bibfnamefont {Y.~V.}\ \bibnamefont {Rostovtsev}}, \ and\
  \bibinfo {author} {\bibfnamefont {M.~O.}\ \bibnamefont {Scully}},\ }\href
  {\doibase 10.1103/PhysRevA.81.063824} {\bibfield  {journal} {\bibinfo
  {journal} {Phys. Rev. A}\ }\textbf {\bibinfo {volume} {81}},\ \bibinfo
  {pages} {063824} (\bibinfo {year} {2010})}\BibitemShut {NoStop}%
\bibitem [{\citenamefont {Harris}(1989)}]{harris}%
  \BibitemOpen
  \bibfield  {author} {\bibinfo {author} {\bibfnamefont {S.~E.}\ \bibnamefont
  {Harris}},\ }\href {\doibase 10.1103/PhysRevLett.62.1033} {\bibfield
  {journal} {\bibinfo  {journal} {Phys. Rev. Lett.}\ }\textbf {\bibinfo
  {volume} {62}},\ \bibinfo {pages} {1033} (\bibinfo {year}
  {1989})}\BibitemShut {NoStop}%
\bibitem [{\citenamefont {Gray}\ \emph {et~al.}(1978)\citenamefont {Gray},
  \citenamefont {Whitley},\ and\ \citenamefont {Stroud}}]{gray}%
  \BibitemOpen
  \bibfield  {author} {\bibinfo {author} {\bibfnamefont {H.~R.}\ \bibnamefont
  {Gray}}, \bibinfo {author} {\bibfnamefont {R.~M.}\ \bibnamefont {Whitley}}, \
  and\ \bibinfo {author} {\bibfnamefont {C.~R.}\ \bibnamefont {Stroud}},\
  }\href {\doibase 10.1364/OL.3.000218} {\bibfield  {journal} {\bibinfo
  {journal} {Opt. Lett.}\ }\textbf {\bibinfo {volume} {3}},\ \bibinfo {pages}
  {218} (\bibinfo {year} {1978})}\BibitemShut {NoStop}%
\bibitem [{\citenamefont {Davuluri}\ and\ \citenamefont
  {Zhu}(2016)}]{sankar-ps}%
  \BibitemOpen
  \bibfield  {author} {\bibinfo {author} {\bibfnamefont {S.}~\bibnamefont
  {Davuluri}}\ and\ \bibinfo {author} {\bibfnamefont {S.}~\bibnamefont {Zhu}},\
  }\href {http://stacks.iop.org/1402-4896/91/i=1/a=013008} {\bibfield
  {journal} {\bibinfo  {journal} {Physica Scripta}\ }\textbf {\bibinfo {volume}
  {91}},\ \bibinfo {pages} {013008} (\bibinfo {year} {2016})}\BibitemShut
  {NoStop}%
\bibitem [{\citenamefont {Fleischhauer}\ \emph {et~al.}(2005)\citenamefont
  {Fleischhauer}, \citenamefont {Imamoglu},\ and\ \citenamefont
  {Marangos}}]{fleischhauer-rmp}%
  \BibitemOpen
  \bibfield  {author} {\bibinfo {author} {\bibfnamefont {M.}~\bibnamefont
  {Fleischhauer}}, \bibinfo {author} {\bibfnamefont {A.}~\bibnamefont
  {Imamoglu}}, \ and\ \bibinfo {author} {\bibfnamefont {J.~P.}\ \bibnamefont
  {Marangos}},\ }\href {\doibase 10.1103/RevModPhys.77.633} {\bibfield
  {journal} {\bibinfo  {journal} {Rev. Mod. Phys.}\ }\textbf {\bibinfo {volume}
  {77}},\ \bibinfo {pages} {633} (\bibinfo {year} {2005})}\BibitemShut
  {NoStop}%
\bibitem [{\citenamefont {Fleischhauer}\ and\ \citenamefont
  {Scully}(1994)}]{fleischhauer-94}%
  \BibitemOpen
  \bibfield  {author} {\bibinfo {author} {\bibfnamefont {M.}~\bibnamefont
  {Fleischhauer}}\ and\ \bibinfo {author} {\bibfnamefont {M.~O.}\ \bibnamefont
  {Scully}},\ }\href {\doibase 10.1103/PhysRevA.49.1973} {\bibfield  {journal}
  {\bibinfo  {journal} {Phys. Rev. A}\ }\textbf {\bibinfo {volume} {49}},\
  \bibinfo {pages} {1973} (\bibinfo {year} {1994})}\BibitemShut {NoStop}%
\bibitem [{\citenamefont {Davuluri}\ and\ \citenamefont
  {Rostovtsev}(2013)}]{sankar-epl1}%
  \BibitemOpen
  \bibfield  {author} {\bibinfo {author} {\bibfnamefont {S.}~\bibnamefont
  {Davuluri}}\ and\ \bibinfo {author} {\bibfnamefont {Y.~V.}\ \bibnamefont
  {Rostovtsev}},\ }\href {http://stacks.iop.org/0295-5075/103/i=2/a=24001}
  {\bibfield  {journal} {\bibinfo  {journal} {EPL (Europhysics Letters)}\
  }\textbf {\bibinfo {volume} {103}},\ \bibinfo {pages} {24001} (\bibinfo
  {year} {2013})}\BibitemShut {NoStop}%
\bibitem [{\citenamefont {Aspelmeyer}\ \emph {et~al.}(2014)\citenamefont
  {Aspelmeyer}, \citenamefont {Kippenberg},\ and\ \citenamefont
  {Marquardt}}]{aspelmeyer-rmp}%
  \BibitemOpen
  \bibfield  {author} {\bibinfo {author} {\bibfnamefont {M.}~\bibnamefont
  {Aspelmeyer}}, \bibinfo {author} {\bibfnamefont {T.~J.}\ \bibnamefont
  {Kippenberg}}, \ and\ \bibinfo {author} {\bibfnamefont {F.}~\bibnamefont
  {Marquardt}},\ }\href {\doibase 10.1103/RevModPhys.86.1391} {\bibfield
  {journal} {\bibinfo  {journal} {Rev. Mod. Phys.}\ }\textbf {\bibinfo {volume}
  {86}},\ \bibinfo {pages} {1391} (\bibinfo {year} {2014})}\BibitemShut
  {NoStop}%
\bibitem [{\citenamefont {Meystre}(2013)}]{meystre}%
  \BibitemOpen
  \bibfield  {author} {\bibinfo {author} {\bibfnamefont {P.}~\bibnamefont
  {Meystre}},\ }\href {\doibase 10.1002/andp.201200226} {\bibfield  {journal}
  {\bibinfo  {journal} {Annalen der Physik}\ }\textbf {\bibinfo {volume}
  {525}},\ \bibinfo {pages} {215} (\bibinfo {year} {2013})}\BibitemShut
  {NoStop}%
\bibitem [{\citenamefont {Norgia}\ and\ \citenamefont {Donati}(2001)}]{norgia}%
  \BibitemOpen
  \bibfield  {author} {\bibinfo {author} {\bibfnamefont {M.}~\bibnamefont
  {Norgia}}\ and\ \bibinfo {author} {\bibfnamefont {S.}~\bibnamefont
  {Donati}},\ }\href {\doibase 10.1049/el:20010520} {\bibfield  {journal}
  {\bibinfo  {journal} {Electronics Letters}\ }\textbf {\bibinfo {volume}
  {37}},\ \bibinfo {pages} {756} (\bibinfo {year} {2001})}\BibitemShut
  {NoStop}%
\bibitem [{\citenamefont {Faust}\ \emph {et~al.}(2013)\citenamefont {Faust},
  \citenamefont {Rieger}, \citenamefont {Seitner}, \citenamefont {Kotthaus},\
  and\ \citenamefont {Weig}}]{faust-nat}%
  \BibitemOpen
  \bibfield  {author} {\bibinfo {author} {\bibfnamefont {T.}~\bibnamefont
  {Faust}}, \bibinfo {author} {\bibfnamefont {J.}~\bibnamefont {Rieger}},
  \bibinfo {author} {\bibfnamefont {M.~J.}\ \bibnamefont {Seitner}}, \bibinfo
  {author} {\bibfnamefont {J.~P.}\ \bibnamefont {Kotthaus}}, \ and\ \bibinfo
  {author} {\bibfnamefont {E.~M.}\ \bibnamefont {Weig}},\ }\href
  {http://dx.doi.org/10.1038/nphys2666} {\bibfield  {journal} {\bibinfo
  {journal} {Nat Phys}\ }\textbf {\bibinfo {volume} {9}},\ \bibinfo {pages}
  {485} (\bibinfo {year} {2013})},\ \bibinfo {note} {letter}\BibitemShut
  {NoStop}%
\bibitem [{\citenamefont {Faust}\ \emph {et~al.}(2012)\citenamefont {Faust},
  \citenamefont {Rieger}, \citenamefont {Seitner}, \citenamefont {Krenn},
  \citenamefont {Kotthaus},\ and\ \citenamefont {Weig}}]{faust-prl}%
  \BibitemOpen
  \bibfield  {author} {\bibinfo {author} {\bibfnamefont {T.}~\bibnamefont
  {Faust}}, \bibinfo {author} {\bibfnamefont {J.}~\bibnamefont {Rieger}},
  \bibinfo {author} {\bibfnamefont {M.~J.}\ \bibnamefont {Seitner}}, \bibinfo
  {author} {\bibfnamefont {P.}~\bibnamefont {Krenn}}, \bibinfo {author}
  {\bibfnamefont {J.~P.}\ \bibnamefont {Kotthaus}}, \ and\ \bibinfo {author}
  {\bibfnamefont {E.~M.}\ \bibnamefont {Weig}},\ }\href {\doibase
  10.1103/PhysRevLett.109.037205} {\bibfield  {journal} {\bibinfo  {journal}
  {Phys. Rev. Lett.}\ }\textbf {\bibinfo {volume} {109}},\ \bibinfo {pages}
  {037205} (\bibinfo {year} {2012})}\BibitemShut {NoStop}%
\bibitem [{\citenamefont {Abram}(1987)}]{abram}%
  \BibitemOpen
  \bibfield  {author} {\bibinfo {author} {\bibfnamefont {I.}~\bibnamefont
  {Abram}},\ }\href {\doibase 10.1103/PhysRevA.35.4661} {\bibfield  {journal}
  {\bibinfo  {journal} {Phys. Rev. A}\ }\textbf {\bibinfo {volume} {35}},\
  \bibinfo {pages} {4661} (\bibinfo {year} {1987})}\BibitemShut {NoStop}%
\bibitem [{\citenamefont {Blow}\ \emph {et~al.}(1990)\citenamefont {Blow},
  \citenamefont {Loudon}, \citenamefont {Phoenix},\ and\ \citenamefont
  {Shepherd}}]{blow}%
  \BibitemOpen
  \bibfield  {author} {\bibinfo {author} {\bibfnamefont {K.~J.}\ \bibnamefont
  {Blow}}, \bibinfo {author} {\bibfnamefont {R.}~\bibnamefont {Loudon}},
  \bibinfo {author} {\bibfnamefont {S.~J.~D.}\ \bibnamefont {Phoenix}}, \ and\
  \bibinfo {author} {\bibfnamefont {T.~J.}\ \bibnamefont {Shepherd}},\ }\href
  {\doibase 10.1103/PhysRevA.42.4102} {\bibfield  {journal} {\bibinfo
  {journal} {Phys. Rev. A}\ }\textbf {\bibinfo {volume} {42}},\ \bibinfo
  {pages} {4102} (\bibinfo {year} {1990})}\BibitemShut {NoStop}%
\bibitem [{\citenamefont {Drummond}\ and\ \citenamefont
  {Carter}(1987)}]{drummond}%
  \BibitemOpen
  \bibfield  {author} {\bibinfo {author} {\bibfnamefont {P.~D.}\ \bibnamefont
  {Drummond}}\ and\ \bibinfo {author} {\bibfnamefont {S.~J.}\ \bibnamefont
  {Carter}},\ }\href {\doibase 10.1364/JOSAB.4.001565} {\bibfield  {journal}
  {\bibinfo  {journal} {J. Opt. Soc. Am. B}\ }\textbf {\bibinfo {volume} {4}},\
  \bibinfo {pages} {1565} (\bibinfo {year} {1987})}\BibitemShut {NoStop}%
\bibitem [{\citenamefont {Fleischhauer}\ and\ \citenamefont
  {Richter}(1995)}]{fleischhauer-95}%
  \BibitemOpen
  \bibfield  {author} {\bibinfo {author} {\bibfnamefont {M.}~\bibnamefont
  {Fleischhauer}}\ and\ \bibinfo {author} {\bibfnamefont {T.}~\bibnamefont
  {Richter}},\ }\href {\doibase 10.1103/PhysRevA.51.2430} {\bibfield  {journal}
  {\bibinfo  {journal} {Phys. Rev. A}\ }\textbf {\bibinfo {volume} {51}},\
  \bibinfo {pages} {2430} (\bibinfo {year} {1995})}\BibitemShut {NoStop}%
\bibitem [{\citenamefont {Das}\ \emph {et~al.}(2005)\citenamefont {Das},
  \citenamefont {Agarwal}, \citenamefont {Golubev},\ and\ \citenamefont
  {Scully}}]{das}%
  \BibitemOpen
  \bibfield  {author} {\bibinfo {author} {\bibfnamefont {K.~K.}\ \bibnamefont
  {Das}}, \bibinfo {author} {\bibfnamefont {G.~S.}\ \bibnamefont {Agarwal}},
  \bibinfo {author} {\bibfnamefont {Y.~M.}\ \bibnamefont {Golubev}}, \ and\
  \bibinfo {author} {\bibfnamefont {M.~O.}\ \bibnamefont {Scully}},\ }\href
  {\doibase 10.1103/PhysRevA.71.013802} {\bibfield  {journal} {\bibinfo
  {journal} {Phys. Rev. A}\ }\textbf {\bibinfo {volume} {71}},\ \bibinfo
  {pages} {013802} (\bibinfo {year} {2005})}\BibitemShut {NoStop}%
\bibitem [{\citenamefont {Dantan}\ and\ \citenamefont
  {Pinard}(2004)}]{dantan-04}%
  \BibitemOpen
  \bibfield  {author} {\bibinfo {author} {\bibfnamefont {A.}~\bibnamefont
  {Dantan}}\ and\ \bibinfo {author} {\bibfnamefont {M.}~\bibnamefont
  {Pinard}},\ }\href {\doibase 10.1103/PhysRevA.69.043810} {\bibfield
  {journal} {\bibinfo  {journal} {Phys. Rev. A}\ }\textbf {\bibinfo {volume}
  {69}},\ \bibinfo {pages} {043810} (\bibinfo {year} {2004})}\BibitemShut
  {NoStop}%
\bibitem [{\citenamefont {Sun}\ \emph {et~al.}(2003)\citenamefont {Sun},
  \citenamefont {Li},\ and\ \citenamefont {Liu}}]{sun}%
  \BibitemOpen
  \bibfield  {author} {\bibinfo {author} {\bibfnamefont {C.~P.}\ \bibnamefont
  {Sun}}, \bibinfo {author} {\bibfnamefont {Y.}~\bibnamefont {Li}}, \ and\
  \bibinfo {author} {\bibfnamefont {X.~F.}\ \bibnamefont {Liu}},\ }\href
  {\doibase 10.1103/PhysRevLett.91.147903} {\bibfield  {journal} {\bibinfo
  {journal} {Phys. Rev. Lett.}\ }\textbf {\bibinfo {volume} {91}},\ \bibinfo
  {pages} {147903} (\bibinfo {year} {2003})}\BibitemShut {NoStop}%
\bibitem [{\citenamefont {Scully}\ and\ \citenamefont
  {Zubairy}(1997)}]{marlan-text}%
  \BibitemOpen
  \bibfield  {author} {\bibinfo {author} {\bibfnamefont {M.~O.}\ \bibnamefont
  {Scully}}\ and\ \bibinfo {author} {\bibfnamefont {S.}~\bibnamefont
  {Zubairy}},\ }\href@noop {} {\emph {\bibinfo {title} {Quantum Optics}}},\
  \bibinfo {edition} {1st}\ ed.\ (\bibinfo  {publisher} {Cambridge University
  Press},\ \bibinfo {year} {1997})\BibitemShut {NoStop}%
\bibitem [{\citenamefont {Novikova}\ \emph {et~al.}(2001)\citenamefont
  {Novikova}, \citenamefont {Matsko}, \citenamefont {Velichansky},
  \citenamefont {Scully},\ and\ \citenamefont {Welch}}]{fleischhauer-00}%
  \BibitemOpen
  \bibfield  {author} {\bibinfo {author} {\bibfnamefont {I.}~\bibnamefont
  {Novikova}}, \bibinfo {author} {\bibfnamefont {A.~B.}\ \bibnamefont
  {Matsko}}, \bibinfo {author} {\bibfnamefont {V.~L.}\ \bibnamefont
  {Velichansky}}, \bibinfo {author} {\bibfnamefont {M.~O.}\ \bibnamefont
  {Scully}}, \ and\ \bibinfo {author} {\bibfnamefont {G.~R.}\ \bibnamefont
  {Welch}},\ }\href {\doibase 10.1103/PhysRevA.63.063802} {\bibfield  {journal}
  {\bibinfo  {journal} {Phys. Rev. A}\ }\textbf {\bibinfo {volume} {63}},\
  \bibinfo {pages} {063802} (\bibinfo {year} {2001})}\BibitemShut {NoStop}%
\bibitem [{\citenamefont {Caves}(1981)}]{caves}%
  \BibitemOpen
  \bibfield  {author} {\bibinfo {author} {\bibfnamefont {C.~M.}\ \bibnamefont
  {Caves}},\ }\href {\doibase 10.1103/PhysRevD.23.1693} {\bibfield  {journal}
  {\bibinfo  {journal} {Phys. Rev. D}\ }\textbf {\bibinfo {volume} {23}},\
  \bibinfo {pages} {1693} (\bibinfo {year} {1981})}\BibitemShut {NoStop}%
\bibitem [{\citenamefont {Carusotto}\ \emph {et~al.}(2003)\citenamefont
  {Carusotto}, \citenamefont {Artoni}, \citenamefont {La~Rocca},\ and\
  \citenamefont {Bassani}}]{carusotto}%
  \BibitemOpen
  \bibfield  {author} {\bibinfo {author} {\bibfnamefont {I.}~\bibnamefont
  {Carusotto}}, \bibinfo {author} {\bibfnamefont {M.}~\bibnamefont {Artoni}},
  \bibinfo {author} {\bibfnamefont {G.~C.}\ \bibnamefont {La~Rocca}}, \ and\
  \bibinfo {author} {\bibfnamefont {F.}~\bibnamefont {Bassani}},\ }\href
  {\doibase 10.1103/PhysRevA.68.063819} {\bibfield  {journal} {\bibinfo
  {journal} {Phys. Rev. A}\ }\textbf {\bibinfo {volume} {68}},\ \bibinfo
  {pages} {063819} (\bibinfo {year} {2003})}\BibitemShut {NoStop}%
\bibitem [{\citenamefont {Davuluri}\ and\ \citenamefont
  {Rostovtsev}(2012)}]{sankar-pra1}%
  \BibitemOpen
  \bibfield  {author} {\bibinfo {author} {\bibfnamefont {S.}~\bibnamefont
  {Davuluri}}\ and\ \bibinfo {author} {\bibfnamefont {Y.~V.}\ \bibnamefont
  {Rostovtsev}},\ }\href {\doibase 10.1103/PhysRevA.86.013806} {\bibfield
  {journal} {\bibinfo  {journal} {Phys. Rev. A}\ }\textbf {\bibinfo {volume}
  {86}},\ \bibinfo {pages} {013806} (\bibinfo {year} {2012})}\BibitemShut
  {NoStop}%
\bibitem [{\citenamefont {Safari}\ \emph {et~al.}(2016)\citenamefont {Safari},
  \citenamefont {De~Leon}, \citenamefont {Mirhosseini}, \citenamefont {Maga\~na
  Loaiza},\ and\ \citenamefont {Boyd}}]{safari}%
  \BibitemOpen
  \bibfield  {author} {\bibinfo {author} {\bibfnamefont {A.}~\bibnamefont
  {Safari}}, \bibinfo {author} {\bibfnamefont {I.}~\bibnamefont {De~Leon}},
  \bibinfo {author} {\bibfnamefont {M.}~\bibnamefont {Mirhosseini}}, \bibinfo
  {author} {\bibfnamefont {O.~S.}\ \bibnamefont {Maga\~na Loaiza}}, \ and\
  \bibinfo {author} {\bibfnamefont {R.~W.}\ \bibnamefont {Boyd}},\ }\href
  {\doibase 10.1103/PhysRevLett.116.013601} {\bibfield  {journal} {\bibinfo
  {journal} {Phys. Rev. Lett.}\ }\textbf {\bibinfo {volume} {116}},\ \bibinfo
  {pages} {013601} (\bibinfo {year} {2016})}\BibitemShut {NoStop}%
\bibitem [{\citenamefont {Post}(1967)}]{post}%
  \BibitemOpen
  \bibfield  {author} {\bibinfo {author} {\bibfnamefont {E.~J.}\ \bibnamefont
  {Post}},\ }\href {\doibase 10.1103/RevModPhys.39.475} {\bibfield  {journal}
  {\bibinfo  {journal} {Rev. Mod. Phys.}\ }\textbf {\bibinfo {volume} {39}},\
  \bibinfo {pages} {475} (\bibinfo {year} {1967})}\BibitemShut {NoStop}%
\bibitem [{\citenamefont {Malykin}(1997)}]{malykin-ss}%
  \BibitemOpen
  \bibfield  {author} {\bibinfo {author} {\bibfnamefont {G.~B.}\ \bibnamefont
  {Malykin}},\ }\href {http://stacks.iop.org/1063-7869/40/i=3/a=R07} {\bibfield
   {journal} {\bibinfo  {journal} {Physics-Uspekhi}\ }\textbf {\bibinfo
  {volume} {40}},\ \bibinfo {pages} {317} (\bibinfo {year} {1997})}\BibitemShut
  {NoStop}%
\bibitem [{\citenamefont {Faucheux}\ \emph {et~al.}(1988)\citenamefont
  {Faucheux}, \citenamefont {Fayoux},\ and\ \citenamefont {Roland}}]{faucheux}%
  \BibitemOpen
  \bibfield  {author} {\bibinfo {author} {\bibfnamefont {M.}~\bibnamefont
  {Faucheux}}, \bibinfo {author} {\bibfnamefont {D.}~\bibnamefont {Fayoux}}, \
  and\ \bibinfo {author} {\bibfnamefont {J.~J.}\ \bibnamefont {Roland}},\
  }\href {http://stacks.iop.org/0150-536X/19/i=3/a=001} {\bibfield  {journal}
  {\bibinfo  {journal} {Journal of Optics}\ }\textbf {\bibinfo {volume} {19}},\
  \bibinfo {pages} {101} (\bibinfo {year} {1988})}\BibitemShut {NoStop}%
\bibitem [{\citenamefont {Cresser}\ \emph
  {et~al.}(1982{\natexlab{a}})\citenamefont {Cresser}, \citenamefont
  {Louisell}, \citenamefont {Meystre}, \citenamefont {Schleich},\ and\
  \citenamefont {Scully}}]{cresser}%
  \BibitemOpen
  \bibfield  {author} {\bibinfo {author} {\bibfnamefont {J.~D.}\ \bibnamefont
  {Cresser}}, \bibinfo {author} {\bibfnamefont {W.~H.}\ \bibnamefont
  {Louisell}}, \bibinfo {author} {\bibfnamefont {P.}~\bibnamefont {Meystre}},
  \bibinfo {author} {\bibfnamefont {W.}~\bibnamefont {Schleich}}, \ and\
  \bibinfo {author} {\bibfnamefont {M.~O.}\ \bibnamefont {Scully}},\ }\href
  {\doibase 10.1103/PhysRevA.25.2214} {\bibfield  {journal} {\bibinfo
  {journal} {Phys. Rev. A}\ }\textbf {\bibinfo {volume} {25}},\ \bibinfo
  {pages} {2214} (\bibinfo {year} {1982}{\natexlab{a}})}\BibitemShut {NoStop}%
\bibitem [{\citenamefont {Schleich}\ \emph {et~al.}(1984)\citenamefont
  {Schleich}, \citenamefont {Cha},\ and\ \citenamefont {Cresser}}]{schleich}%
  \BibitemOpen
  \bibfield  {author} {\bibinfo {author} {\bibfnamefont {W.}~\bibnamefont
  {Schleich}}, \bibinfo {author} {\bibfnamefont {C.~S.}\ \bibnamefont {Cha}}, \
  and\ \bibinfo {author} {\bibfnamefont {J.~D.}\ \bibnamefont {Cresser}},\
  }\href {\doibase 10.1103/PhysRevA.29.230} {\bibfield  {journal} {\bibinfo
  {journal} {Phys. Rev. A}\ }\textbf {\bibinfo {volume} {29}},\ \bibinfo
  {pages} {230} (\bibinfo {year} {1984})}\BibitemShut {NoStop}%
\bibitem [{\citenamefont {Cresser}\ \emph
  {et~al.}(1982{\natexlab{b}})\citenamefont {Cresser}, \citenamefont
  {Hammonds}, \citenamefont {Louisell}, \citenamefont {Meystre},\ and\
  \citenamefont {Risken}}]{cresser-2}%
  \BibitemOpen
  \bibfield  {author} {\bibinfo {author} {\bibfnamefont {J.~D.}\ \bibnamefont
  {Cresser}}, \bibinfo {author} {\bibfnamefont {D.}~\bibnamefont {Hammonds}},
  \bibinfo {author} {\bibfnamefont {W.~H.}\ \bibnamefont {Louisell}}, \bibinfo
  {author} {\bibfnamefont {P.}~\bibnamefont {Meystre}}, \ and\ \bibinfo
  {author} {\bibfnamefont {H.}~\bibnamefont {Risken}},\ }\href {\doibase
  10.1103/PhysRevA.25.2226} {\bibfield  {journal} {\bibinfo  {journal} {Phys.
  Rev. A}\ }\textbf {\bibinfo {volume} {25}},\ \bibinfo {pages} {2226}
  (\bibinfo {year} {1982}{\natexlab{b}})}\BibitemShut {NoStop}%
\bibitem [{\citenamefont {Lef{\`e}vre}(1997)}]{lefevre}%
  \BibitemOpen
  \bibfield  {author} {\bibinfo {author} {\bibfnamefont {H.~C.}\ \bibnamefont
  {Lef{\`e}vre}},\ }\href {\doibase 10.1007/BF02935984} {\bibfield  {journal}
  {\bibinfo  {journal} {Optical Review}\ }\textbf {\bibinfo {volume} {4}},\
  \bibinfo {pages} {A20} (\bibinfo {year} {1997})}\BibitemShut {NoStop}%
\bibitem [{\citenamefont {Vali}\ and\ \citenamefont {Shorthill}(1976)}]{vali}%
  \BibitemOpen
  \bibfield  {author} {\bibinfo {author} {\bibfnamefont {V.}~\bibnamefont
  {Vali}}\ and\ \bibinfo {author} {\bibfnamefont {R.~W.}\ \bibnamefont
  {Shorthill}},\ }\href {\doibase 10.1364/AO.15.001099} {\bibfield  {journal}
  {\bibinfo  {journal} {Appl. Opt.}\ }\textbf {\bibinfo {volume} {15}},\
  \bibinfo {pages} {1099} (\bibinfo {year} {1976})}\BibitemShut {NoStop}%
\bibitem [{\citenamefont {Schiller}(2013)}]{schiller}%
  \BibitemOpen
  \bibfield  {author} {\bibinfo {author} {\bibfnamefont {S.}~\bibnamefont
  {Schiller}},\ }\href {\doibase 10.1103/PhysRevA.87.033823} {\bibfield
  {journal} {\bibinfo  {journal} {Phys. Rev. A}\ }\textbf {\bibinfo {volume}
  {87}},\ \bibinfo {pages} {033823} (\bibinfo {year} {2013})}\BibitemShut
  {NoStop}%
\bibitem [{\citenamefont {Scully}\ and\ \citenamefont
  {Dowling}(1993)}]{marlan-mwg}%
  \BibitemOpen
  \bibfield  {author} {\bibinfo {author} {\bibfnamefont {M.~O.}\ \bibnamefont
  {Scully}}\ and\ \bibinfo {author} {\bibfnamefont {J.~P.}\ \bibnamefont
  {Dowling}},\ }\href {\doibase 10.1103/PhysRevA.48.3186} {\bibfield  {journal}
  {\bibinfo  {journal} {Phys. Rev. A}\ }\textbf {\bibinfo {volume} {48}},\
  \bibinfo {pages} {3186} (\bibinfo {year} {1993})}\BibitemShut {NoStop}%
\bibitem [{\citenamefont {Campbell}\ and\ \citenamefont
  {Hamilton}(2017)}]{campbell}%
  \BibitemOpen
  \bibfield  {author} {\bibinfo {author} {\bibfnamefont {W.~C.}\ \bibnamefont
  {Campbell}}\ and\ \bibinfo {author} {\bibfnamefont {P.}~\bibnamefont
  {Hamilton}},\ }\href {http://stacks.iop.org/0953-4075/50/i=6/a=064002}
  {\bibfield  {journal} {\bibinfo  {journal} {Journal of Physics B: Atomic,
  Molecular and Optical Physics}\ }\textbf {\bibinfo {volume} {50}},\ \bibinfo
  {pages} {064002} (\bibinfo {year} {2017})}\BibitemShut {NoStop}%
\bibitem [{\citenamefont {Wu}\ \emph {et~al.}(2007)\citenamefont {Wu},
  \citenamefont {Su},\ and\ \citenamefont {Prentiss}}]{wu}%
  \BibitemOpen
  \bibfield  {author} {\bibinfo {author} {\bibfnamefont {S.}~\bibnamefont
  {Wu}}, \bibinfo {author} {\bibfnamefont {E.}~\bibnamefont {Su}}, \ and\
  \bibinfo {author} {\bibfnamefont {M.}~\bibnamefont {Prentiss}},\ }\href
  {\doibase 10.1103/PhysRevLett.99.173201} {\bibfield  {journal} {\bibinfo
  {journal} {Phys. Rev. Lett.}\ }\textbf {\bibinfo {volume} {99}},\ \bibinfo
  {pages} {173201} (\bibinfo {year} {2007})}\BibitemShut {NoStop}%
\bibitem [{\citenamefont {Qi}\ \emph {et~al.}(2017)\citenamefont {Qi},
  \citenamefont {Hu}, \citenamefont {Valenzuela}, \citenamefont {Zhang},
  \citenamefont {Zhai}, \citenamefont {Quan}, \citenamefont {Waltham},\ and\
  \citenamefont {Fang}}]{qi}%
  \BibitemOpen
  \bibfield  {author} {\bibinfo {author} {\bibfnamefont {L.}~\bibnamefont
  {Qi}}, \bibinfo {author} {\bibfnamefont {Z.}~\bibnamefont {Hu}}, \bibinfo
  {author} {\bibfnamefont {T.}~\bibnamefont {Valenzuela}}, \bibinfo {author}
  {\bibfnamefont {Y.}~\bibnamefont {Zhang}}, \bibinfo {author} {\bibfnamefont
  {Y.}~\bibnamefont {Zhai}}, \bibinfo {author} {\bibfnamefont {W.}~\bibnamefont
  {Quan}}, \bibinfo {author} {\bibfnamefont {N.}~\bibnamefont {Waltham}}, \
  and\ \bibinfo {author} {\bibfnamefont {J.}~\bibnamefont {Fang}},\ }\href
  {\doibase 10.1063/1.4980066} {\bibfield  {journal} {\bibinfo  {journal}
  {Applied Physics Letters}\ }\textbf {\bibinfo {volume} {110}},\ \bibinfo
  {pages} {153502} (\bibinfo {year} {2017})},\ \Eprint
  {http://arxiv.org/abs/https://doi.org/10.1063/1.4980066}
  {https://doi.org/10.1063/1.4980066} \BibitemShut {NoStop}%
\bibitem [{\citenamefont {Hurst}\ \emph {et~al.}(2007)\citenamefont {Hurst},
  \citenamefont {Wells},\ and\ \citenamefont {Stedman}}]{hurst}%
  \BibitemOpen
  \bibfield  {author} {\bibinfo {author} {\bibfnamefont {R.~B.}\ \bibnamefont
  {Hurst}}, \bibinfo {author} {\bibfnamefont {J.-P.~R.}\ \bibnamefont {Wells}},
  \ and\ \bibinfo {author} {\bibfnamefont {G.~E.}\ \bibnamefont {Stedman}},\
  }\href {http://stacks.iop.org/1464-4258/9/i=10/a=010} {\bibfield  {journal}
  {\bibinfo  {journal} {Journal of Optics A: Pure and Applied Optics}\ }\textbf
  {\bibinfo {volume} {9}},\ \bibinfo {pages} {838} (\bibinfo {year}
  {2007})}\BibitemShut {NoStop}%
\bibitem [{\citenamefont {Caves}(1980)}]{caves-rp}%
  \BibitemOpen
  \bibfield  {author} {\bibinfo {author} {\bibfnamefont {C.~M.}\ \bibnamefont
  {Caves}},\ }\href {\doibase 10.1103/PhysRevLett.45.75} {\bibfield  {journal}
  {\bibinfo  {journal} {Phys. Rev. Lett.}\ }\textbf {\bibinfo {volume} {45}},\
  \bibinfo {pages} {75} (\bibinfo {year} {1980})}\BibitemShut {NoStop}%
\bibitem [{\citenamefont {Hänsch}\ and\ \citenamefont
  {Schawlow}(1975)}]{hansch}%
  \BibitemOpen
  \bibfield  {author} {\bibinfo {author} {\bibfnamefont {T.}~\bibnamefont
  {Hänsch}}\ and\ \bibinfo {author} {\bibfnamefont {A.}~\bibnamefont
  {Schawlow}},\ }\href {\doibase https://doi.org/10.1016/0030-4018(75)90159-5}
  {\bibfield  {journal} {\bibinfo  {journal} {Optics Communications}\ }\textbf
  {\bibinfo {volume} {13}},\ \bibinfo {pages} {68 } (\bibinfo {year}
  {1975})}\BibitemShut {NoStop}%
\bibitem [{\citenamefont {Wineland}\ \emph {et~al.}(1978)\citenamefont
  {Wineland}, \citenamefont {Drullinger},\ and\ \citenamefont
  {Walls}}]{wineland}%
  \BibitemOpen
  \bibfield  {author} {\bibinfo {author} {\bibfnamefont {D.~J.}\ \bibnamefont
  {Wineland}}, \bibinfo {author} {\bibfnamefont {R.~E.}\ \bibnamefont
  {Drullinger}}, \ and\ \bibinfo {author} {\bibfnamefont {F.~L.}\ \bibnamefont
  {Walls}},\ }\href {\doibase 10.1103/PhysRevLett.40.1639} {\bibfield
  {journal} {\bibinfo  {journal} {Phys. Rev. Lett.}\ }\textbf {\bibinfo
  {volume} {40}},\ \bibinfo {pages} {1639} (\bibinfo {year}
  {1978})}\BibitemShut {NoStop}%
\bibitem [{\citenamefont {Arcizet}\ \emph {et~al.}(2006)\citenamefont
  {Arcizet}, \citenamefont {Cohadon}, \citenamefont {Briant}, \citenamefont
  {Pinard}, \citenamefont {Heidmann}, \citenamefont {Mackowski}, \citenamefont
  {Michel}, \citenamefont {Pinard}, \citenamefont
  {Fran\ifmmode~\mbox{\c{c}}\else \c{c}\fi{}ais},\ and\ \citenamefont
  {Rousseau}}]{arcizet}%
  \BibitemOpen
  \bibfield  {author} {\bibinfo {author} {\bibfnamefont {O.}~\bibnamefont
  {Arcizet}}, \bibinfo {author} {\bibfnamefont {P.-F.}\ \bibnamefont
  {Cohadon}}, \bibinfo {author} {\bibfnamefont {T.}~\bibnamefont {Briant}},
  \bibinfo {author} {\bibfnamefont {M.}~\bibnamefont {Pinard}}, \bibinfo
  {author} {\bibfnamefont {A.}~\bibnamefont {Heidmann}}, \bibinfo {author}
  {\bibfnamefont {J.-M.}\ \bibnamefont {Mackowski}}, \bibinfo {author}
  {\bibfnamefont {C.}~\bibnamefont {Michel}}, \bibinfo {author} {\bibfnamefont
  {L.}~\bibnamefont {Pinard}}, \bibinfo {author} {\bibfnamefont
  {O.}~\bibnamefont {Fran\ifmmode~\mbox{\c{c}}\else \c{c}\fi{}ais}}, \ and\
  \bibinfo {author} {\bibfnamefont {L.}~\bibnamefont {Rousseau}},\ }\href
  {\doibase 10.1103/PhysRevLett.97.133601} {\bibfield  {journal} {\bibinfo
  {journal} {Phys. Rev. Lett.}\ }\textbf {\bibinfo {volume} {97}},\ \bibinfo
  {pages} {133601} (\bibinfo {year} {2006})}\BibitemShut {NoStop}%
\bibitem [{\citenamefont {Marquardt}\ \emph {et~al.}(2007)\citenamefont
  {Marquardt}, \citenamefont {Chen}, \citenamefont {Clerk},\ and\ \citenamefont
  {Girvin}}]{marquardt}%
  \BibitemOpen
  \bibfield  {author} {\bibinfo {author} {\bibfnamefont {F.}~\bibnamefont
  {Marquardt}}, \bibinfo {author} {\bibfnamefont {J.~P.}\ \bibnamefont {Chen}},
  \bibinfo {author} {\bibfnamefont {A.~A.}\ \bibnamefont {Clerk}}, \ and\
  \bibinfo {author} {\bibfnamefont {S.~M.}\ \bibnamefont {Girvin}},\ }\href
  {\doibase 10.1103/PhysRevLett.99.093902} {\bibfield  {journal} {\bibinfo
  {journal} {Phys. Rev. Lett.}\ }\textbf {\bibinfo {volume} {99}},\ \bibinfo
  {pages} {093902} (\bibinfo {year} {2007})}\BibitemShut {NoStop}%
\bibitem [{\citenamefont {Bhattacharya}\ and\ \citenamefont
  {Meystre}(2007)}]{bhattacharya}%
  \BibitemOpen
  \bibfield  {author} {\bibinfo {author} {\bibfnamefont {M.}~\bibnamefont
  {Bhattacharya}}\ and\ \bibinfo {author} {\bibfnamefont {P.}~\bibnamefont
  {Meystre}},\ }\href {\doibase 10.1103/PhysRevLett.99.073601} {\bibfield
  {journal} {\bibinfo  {journal} {Phys. Rev. Lett.}\ }\textbf {\bibinfo
  {volume} {99}},\ \bibinfo {pages} {073601} (\bibinfo {year}
  {2007})}\BibitemShut {NoStop}%
\bibitem [{\citenamefont {Schliesser}\ \emph {et~al.}(2008)\citenamefont
  {Schliesser}, \citenamefont {Riviere}, \citenamefont {Anetsberger},
  \citenamefont {Arcizet},\ and\ \citenamefont
  {Kippenberg}}]{schliesser-cooling}%
  \BibitemOpen
  \bibfield  {author} {\bibinfo {author} {\bibfnamefont {A.}~\bibnamefont
  {Schliesser}}, \bibinfo {author} {\bibfnamefont {R.}~\bibnamefont {Riviere}},
  \bibinfo {author} {\bibfnamefont {G.}~\bibnamefont {Anetsberger}}, \bibinfo
  {author} {\bibfnamefont {O.}~\bibnamefont {Arcizet}}, \ and\ \bibinfo
  {author} {\bibfnamefont {T.~J.}\ \bibnamefont {Kippenberg}},\ }\href
  {http://dx.doi.org/10.1038/nphys939} {\bibfield  {journal} {\bibinfo
  {journal} {Nat. Phys.}\ }\textbf {\bibinfo {volume} {4}},\ \bibinfo {pages}
  {415} (\bibinfo {year} {2008})}\BibitemShut {NoStop}%
\end{thebibliography}%
\end{document}